\documentclass[aps,pre,superscriptaddress,a4paper,floatfix,showpacs]{revtex4}
\usepackage{graphicx,amsmath,amssymb,psfrag,curves,epic,subeqnarray}
\DeclareMathAlphabet{\mathpzc}{OT1}{pzc}{m}{it}
\newcommand{\bu}{\mathbf u} 
\newcommand{\bU}{\mathbf U} 

\def\tdel{\Delta t}
\def\lap{\nabla^2}
\def\Usol{\mathcal{U}}
\def\usol{\mathpzc{u}}
\def\Hmax{\bar{H}}
\def\etal{{\em et al.}}
\begin{document}

\title{Extreme multiplicity in cylindrical Rayleigh-B\'enard convection:\\
II.~Bifurcation diagram and symmetry classification}

\author{Katarzyna Boro\'nska}
\email[]{k.boronska@leeds.ac.uk}
\homepage[]{www.comp.leeds.ac.uk/kb}
\affiliation{School of Computing, University of Leeds, 
Leeds LS2 9JT, United Kingdom}

\author{Laurette S.\ Tuckerman}
\email[]{laurette@pmmh.espci.fr}
\homepage[]{www.pmmh.espci.fr/~laurette}
\affiliation{PMMH-ESPCI, CNRS (UMR7636), Univ. Paris VI \& VII,
10 rue Vauquelin, 75231 Paris France}

\begin{abstract}
A large number of flows with distinctive patterns have been observed in
experiments and simulations of Rayleigh-B\'enard convection in a water-filled
cylinder whose radius is twice the height. We have adapted a time-dependent
pseudospectral code, first, to carry out Newton's method and branch
continuation and, second, to carry out the exponential power method and
Arnoldi iteration to calculate leading eigenpairs and determine the stability
of the steady states.  The resulting bifurcation diagram represents a
compromise between the tendency in the bulk towards parallel rolls, and the
requirement imposed by the boundary conditions that primary bifurcations be
towards states whose azimuthal dependence is trigonometric.  The diagram
contains 17 branches of stable and unstable steady states. These can be
classified geometrically as roll states containing two, three, and four rolls;
axisymmetric patterns with one or two tori; three-fold symmetric patterns
called mercedes, mitubishi, marigold and cloverleaf; trigonometric patterns
called dipole and pizza; and less symmetric patterns called CO and asymmetric
three-rolls.  The convective branches are connected to the conductive state
and to each other by 16 primary and secondary pitchfork bifurcations and
turning points.  In order to better understand this complicated bifurcation
diagram, we have partitioned it according to azimuthal symmetry.  We have been
able to determine the bifurcation-theoretic origin from the conductive state
of all the branches observed at high Rayleigh number.
\end{abstract}
\date{\today}
\maketitle

\section{Introduction}

\begin{figure}[htp]
\includegraphics[width=17cm]{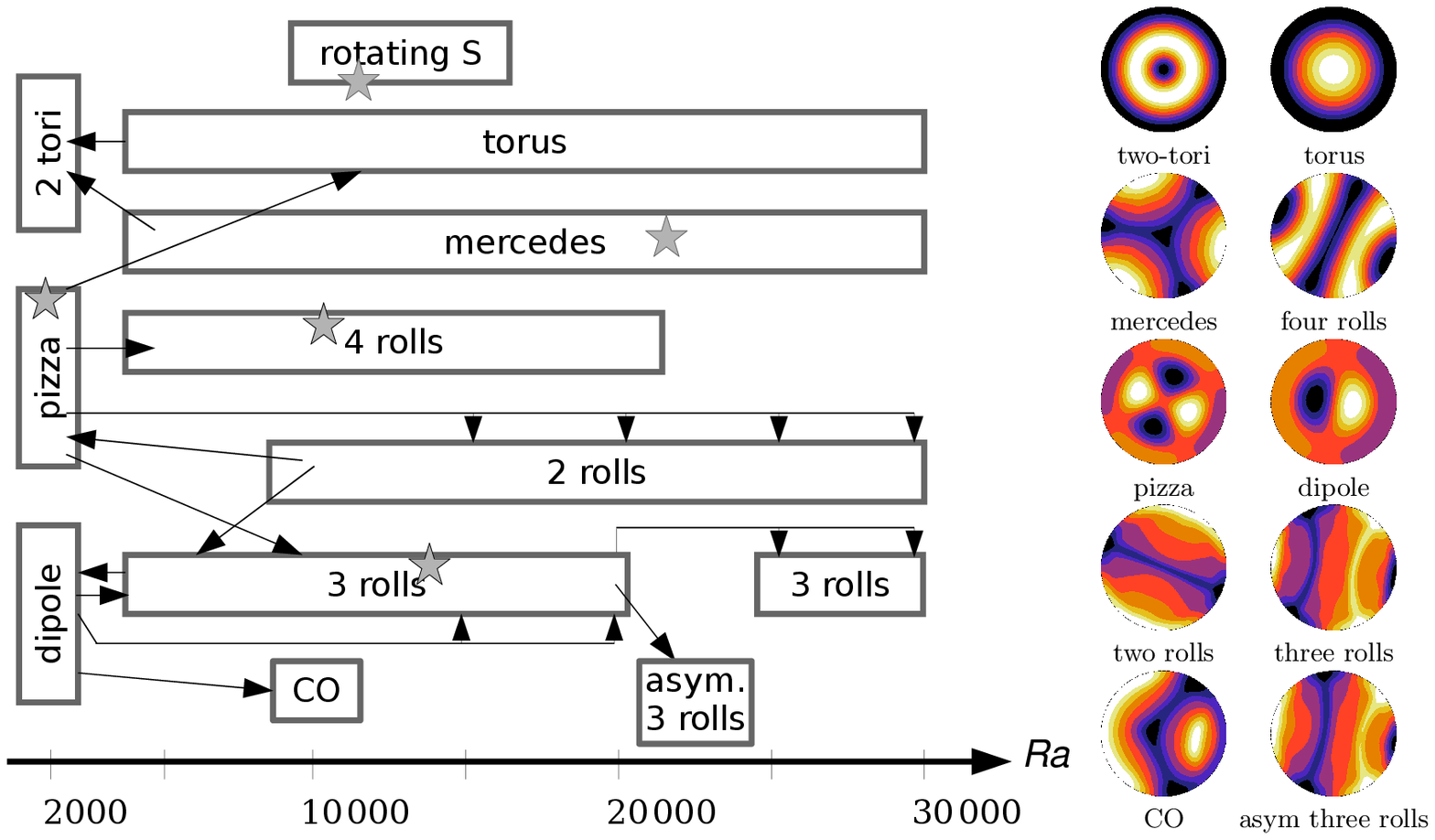}
\caption{(Color online) Schematic diagram of existence ranges and transitions between
  convective patterns observed in time-dependent simulation for insulating
  sidewalls.  Stars denote solutions obtained from a slight perturbation of
  the conductive state, at the Rayleigh numbers indicated.  The initial
  condition was identical for all five simulations.  Arrows indicate patterns
  obtained by starting from stable steady states, and abruptly either lowering
  or raising the Rayleigh number.  For example, at $Ra=2000$, the perturbed
  conductive initial condition leads to a pizza state. Using the pizza state
  at $Ra=2000$ as an initial condition leads to a four-roll state at
  $Ra=5000$, to the three-roll state at $Ra=10 \,000$, and to a two-roll state
  at $Ra\geq 15 \,000$. \\ Right: representative patterns illustrated via
  temperature field in the horizontal midplane, with light portions
  representing hot rising fluid and dark portions representing cold descending
  fluid.}
\label{fig:dns:neu}
\end{figure}

In the late 1990s, Hof, Lucas and Mullin~\cite{Hof,HofThesis} described five distinct
steady patterns observed experimentally in a cylindrical Rayleigh-B\'enard
convection cell at identical parameter values.  More precisely, the patterns
observed were torus, two-, three-, and four-roll states, and a mercedes
pattern, at Prandtl number $Pr=6.7$, Rayleigh number $Ra=14\,200$, and an
aspect ratio $\Gamma\equiv$radius/height=2 with insulating lateral boundaries.
In our previous work~\cite{Boronska_PhD,Boronska_PRE1}, we reproduced
numerically the five patterns observed by Hof and determined the approximate
limits in Rayleigh number over which they could be observed.  At lower
Rayleigh numbers, we simulated several other patterns -- dipole, pizza,
and two-tori -- as well as some time-periodic patterns. These results are
summarized in figure \ref{fig:dns:neu}.
%
Our viewpoint, pioneered in the 1980s by Benjamin and Mullin~\cite{Benjamin},
is that these observations can be best understood and organized by 
constructing the bifurcation diagram corresponding to this figure.
In particular, we wish to trace connections between the patterns 
observed at high and at low Rayleigh numbers, and to 
the basic conductive state wherever possible.

The classical analysis of onset of Rayleigh-B\'enard convection describes an
instability of the conductive state to a pattern of straight parallel rolls of
infinite length.  However, such a pattern is clearly not realizable in a
small-aspect-ratio cylinder. Rolls must be curved to fit into the container,
as shown in the two-, three- and four-roll states illustrated in figure
\ref{fig:dns:neu}.  In addition, a primary bifurcation, that is, a bifurcation
from the conductive state, is associated with an eigenmode which is
necessarily trigonometric in the azimuthal angle, such as the dipole or pizza
states, or two-tori and torus states of figure \ref{fig:dns:neu}.  The focus
of this paper is the relationship between trigonometric modes and roll states
and, more generally, the bifurcation-theoretic genesis of the profusion of
states in this configuration.

In our companion paper~\cite{Boronska_PRE1}, we reviewed some of the
literature on Rayleigh-B\'enard convection in small-aspect-ratio cylindrical
geometries, focusing on pattern competition.  The previous investigations most
relevant to this manuscript, in addition to those of Hof et
al.~\cite{Hof,HofThesis}, are the full nonlinear simulations of Leong
\cite{Leong} and of Ma \etal~\cite{Ma}; we will compare our results to these
articles where appropriate.

In section \ref{sec:background} we state the governing equations and the
symmetries of the configuration. Section \ref{sec:numerical} describes the
numerical methods we have used to compute steady states and their stability.
Section \ref{sec:results} begins by presenting the full bifurcation diagram
and primary bifurcations. We then give a detailed analysis of branches
corresponding to each azimuthal wavenumber.  Concluding remarks are presented
in section \ref{sec:conclusion}.

\section{Background}
\label{sec:background}

\subsection{Governing equations and boundary conditions}
\label{sec:equations}

We recall from our companion paper~\cite{Boronska_PRE1} the dimensionless
Navier--Stokes and Boussinesq equations governing the system:
\begin{subequations}
\begin{eqnarray}
\label{eq:N-S}
\partial _t H
+\left(\bU\cdot\nabla\right)H &=& Ra\;U_z+\lap H \label{eq:N-Sb} , \\
Pr^{-1}\left({\partial _t}\bU +\left(\bU\cdot{\mathbf\nabla}\right)\bU\right)
 &=& - {\mathbf\nabla}P + \lap\mathbf{U} +H{\mathbf e_z} \label{eq:N-Sa}, \\
{\mathbf\nabla}\cdot\mathbf{U}&=&0 ,
\label{eq:N-Sc}
\end{eqnarray}
\end{subequations}
where $H$ is the nondimensionalized deviation of the temperature from the 
linear vertical conductive profile.
The parameter values are as follows:
\begin{equation}
Pr =6.7 , \qquad\qquad 
\Gamma \equiv \frac{\rm radius}{\rm height} = 2 , \qquad\qquad
0 \leq Ra \leq 30\,000 .
\end{equation}
The container is assumed to have rigid walls, with thermally 
conducting horizontal bounding plates and thermally insulating sidewalls
\begin{subequations}
\label{eq:bcs}
\begin{eqnarray}
{\bf U} = 0 &&\quad {\rm for} \quad  z=\pm 1/2 \quad {\rm or} 
\quad r=\Gamma , \\ 
	H= 0 &&\quad{\rm for}\quad z=\pm 1/2 , \\
	\partial_r H= 0 &&\quad {\rm for}\quad r=\Gamma
\end{eqnarray}
\end{subequations}

\subsection{Symmetries}
\label{sec:symmetries}

The bifurcations that this system can undergo are dictated by its symmetries.
In group-theoretic terms, the conductive state has $O(2)$ symmetry in the
azimuthal angle, meaning that it is invariant under all rotations and
reflections in $\theta$:
\begin{subequations}
\begin{eqnarray}
(U_r, U_\theta,U_z,H)(r,\theta,z) &=&
(U_r, U_\theta,U_z,H)(r,\theta+\theta_0,z) \label{eq:o2rot}\\
(U_r, U_\theta,U_z,H)(r,\theta,z) &=&
(U_r, -U_\theta,U_z,H)(r,\theta_0-\theta,z) \label{eq:o2ref}
\end{eqnarray}
where $\theta_0$ indicates an arbitrary angle of rotation or axis of
reflection, and all compositions of these transformations.
Under the Boussinesq approximation, the conductive state is also invariant
under simultaneous reflection in $z$ and change in sign of the temperature
perturbation:
\begin{equation*}
(U_r, U_\theta,U_z,H)(r,\theta,z) =
(U_r, U_\theta,-U_z,-H)(r,\theta,-z) 
\end{equation*}
This symmetry can be combined with the 
$\theta$-rotation symmetry \eqref{eq:o2rot} to yield:
\begin{equation}
(U_r, U_\theta,U_z,H)(r,\theta,z) =
(U_r, U_\theta,-U_z,-H)(r,\theta+\theta_0,-z) 
\label{eq:ztempsym}
\end{equation}
\label{eq:o2}
\end{subequations}
a form whose utility will appear shortly.  The full symmetry group of the
conductive state is thus $O(2) \times Z_2$.

A steady bifurcation from the axisymmetric conductive state, i.e.~a primary
bifurcation, is necessarily associated with an eigenvector which is
trigonometric in the azimuthal direction; see, 
e.g. Crawford \& Knobloch~\cite{CK}.  Each bifurcating
branch is thus associated with an azimuthal wavenumber $m$.
For $m=0$, symmetry \eqref{eq:ztempsym} is broken and the bifurcation
is a pitchfork, leading to two branches.  
If $m$ is non-zero, the bifurcation is a circle pitchfork, producing 
families of states of arbitrary orientation.  
For the bifurcating states, $O(2)$ symmetry is replaced by
$D_m$, meaning that they are invariant under rotation by angles which are
multiples of $2\pi/m$ and reflections in $2m$ axes of symmetry:
\begin{subequations}
\begin{eqnarray}
(U_r, U_\theta,U_z,H)(r,\theta,z) &= &
(U_r, U_\theta,U_z,H)(r,\theta+2\pi/m,z) \label{eq:dmrot}\\
(U_r, U_\theta,U_z,H)(r,\theta,z) &= &
(U_r, -U_\theta,U_z,H)(r,-\theta,z) \label{eq:dmref}\\
(U_r, U_\theta,U_z,H)(r,\theta,z) &= & 
(U_r, U_\theta,-U_z,-H)(r,\theta+\pi/m,-z) \label{eq:dmz}
\end{eqnarray}
\label{eq:dm}
\end{subequations}
where $\theta=0$ is taken to be one of the axes of symmetry of the pattern,
and \eqref{eq:dmrot} is trivially verified if $m=1$.
These equations generate the symmetry group $D_m\times Z_2$.  These states
have a zero eigenvalue, corresponding to the marginal stability to rotation of
the pattern.  Equations \eqref{eq:dm} can be seen to be special cases of
\eqref{eq:o2}. (The form of \eqref{eq:dmz} is the reason we choose 
\eqref{eq:ztempsym}, instead of the Boussinesq reflection operator,
 as a generator of the symmetry group.)

Primary branches can undergo secondary pitchfork bifurcations which break the
$Z_2$ symmetry \eqref{eq:dmz}.
The resulting branches, which we will call ``asymmetric'', 
nonetheless have $D_m$ symmetry, 
generated by the discrete rotation symmetry \eqref{eq:dmrot}
and the reflection symmetry \eqref{eq:dmref}.

\section{Numerical Methods}
\label{sec:numerical}

In~\cite{Boronska_PhD,Boronska_PRE1} we described our code for integrating the
time-dependent Boussinesq equations in a cylindrical geometry.
We have modified this time-dependent code using the techniques described in
\cite{Mamun,Timesteppers} to carry out continuation by Newton's method and
linear stability analysis by the exponential Arnoldi method.  We describe
these modifications in the subsections which follow.  To do so, we will write
the Boussinesq equations schematically as
\begin{equation}
\frac{d\:\Usol}{dt} = \mathcal{L}\Usol + \mathcal{N}(\Usol)
\label{eq:schematic}\end{equation}
where $\mathcal{L}$ represents the viscous, diffusive and buoyancy operators
and $\mathcal{N}$ the advective terms.  $\Usol\equiv(H,U_r,U_\theta,U_z)$
represents the spatially discretized temperature deviation $H$ and velocity
field $\bU=(U_r,U_\theta,U_z)$.  The imposition of boundary conditions and
incompressibility are assumed to be included in the representations of
$\mathcal{L}$, $\mathcal{N}$ and $\Usol$.
Here, we assume that timestepping is carried out via the first-order formula:
\begin{equation}
\Usol(t+\Delta t) = (I-\Delta t
\mathcal{L})^{-1}(I + \Delta t \mathcal{N}) \:\Usol(t)
\equiv \mathcal{B}(\Usol(t))
\label{eq:timestep}\end{equation}
i.e.~the terms in $\mathcal{L}$ are treated via the implicit backwards Euler
scheme and those in $\mathcal{N}$ by the explicit forwards Euler scheme.

\subsection{Spatial discretization}
\label{sec:spatial}

The code uses a pseudo-spectral spatial discretization, in which $H$, $U_z$ are
approximated as:
 \begin{equation}
f(r,\theta,z)=\sum^{N_\theta/2}_{m=0}
\sum^{2N_r-1}_{{j\geq m}\atop{j+m\ \mathrm{even}}}
\sum^{N_z-1}_{k=0}\hat{f}_{j,m,k} \: T_j(r/\Gamma)\: T_k(2z) \: e^{im\theta}
+{\rm c.c.}
\label{eq:ztRep2}
\end{equation}
while $j+m$ odd is used for $U_r$, $U_\theta$.  Differentiation is carried out on the
spectral representation \eqref{eq:ztRep2}, while multiplications are performed
after transforming to a grid, and then transforming the result back to the
spectral representation.  
For the aspect ratio $\Gamma=2$
investigated here, we use $N_r=40$ gridpoints or Chebyshev polynomials in the
radial direction, $N_{\theta}=120$ gridpoints or trigonometric functions in
the azimuthal direction and $N_z=20$ gridpoints or Chebyshev polynomials in
the axial direction.  Thus the domain is represented by approximately $10^5$
gridpoints and each solution by a vector of size $4\times 10^5$.  (We have
also checked our resolution for $Ra\geq 20\,000$ by re-calculating a few of
our branches -- the mercedes, one-torus, two-roll and asymmetric three-roll
branches -- with a resolution of $N_r\times N_\theta \times N_z = 60 \times
160 \times 30$.)  The boundary conditions are imposed via the tau method, and
incompressibility to machine accuracy is insured via an influence matrix
technique.

\subsection{Steady state solving}

Steady states are found by calculating the roots of $\mathcal{B}-I$, which are
the same as those of $\mathcal{N}+\mathcal{L}$ {\it for any value of} $\Delta
t$, as shown by the following calculation:
\begin{eqnarray}
(\mathcal{B} - I) &=& (I-\Delta t \mathcal{L})^{-1}(I + \Delta t \mathcal{N}) - I \nonumber\\
&=& (I-\Delta t \mathcal{L})^{-1}\left[(I + \Delta t \mathcal{N}) - (I-\Delta t \mathcal{L})\right]\nonumber\\
&=& (I-\Delta t \mathcal{L})^{-1}\Delta t (\mathcal{N} + \mathcal{L}).\label{eq:difftimestep}
\end{eqnarray}
The roots of $\mathcal{B}-I$ are found by Newton iteration:
\begin{subequations}
\begin{eqnarray}
(\mathcal{B}_\Usol-I) \: \usol &=& (\mathcal{B}-I) \:\Usol \label{eq:BCGS}\\
\Usol &\leftarrow& \Usol-\usol , \label{eq:decrement}
\end{eqnarray}
\label{eq:Newton}
\end{subequations}
where the linear operator $\mathcal{B}_\Usol-I$ is the Jacobian of
$\mathcal{B}-I$ evaluated at $\Usol$:
\begin{equation}(
\mathcal{B}_\Usol - I) \usol= (I-\Delta t \mathcal{L})^{-1}\Delta t
  (\mathcal{N}_\Usol + \mathcal{L}) \usol 
\end{equation}
while $\Usol\equiv(H,\bU)$ is the current estimate for the steady state and
$\usol\equiv(h,\bu)$ is an unknown correction to $\Usol$.  The action
$\mathcal{N}_\Usol \usol$ is obtained from $\mathcal{N}(\Usol)$ merely by
carrying out the replacements
\begin{subequations}
\begin{eqnarray}
\bU\cdot\nabla H &\rightarrow& \bU\cdot\nabla h + \bu\cdot\nabla H \\
\bU\cdot\nabla\bU &\rightarrow& \bU\cdot\nabla\bu + \bu\cdot\nabla\bU
\end{eqnarray}\label{eq:jacobian}\end{subequations}
in the nonlinear terms of \eqref{eq:N-Sa}-\eqref{eq:N-Sb}.
Since the boundary conditions \eqref{eq:bcs} are
homogeneous, they remain unchanged.

We iterate \eqref{eq:Newton} until $||(\mathcal{B}-I)\:\Usol||$ is lower than
some threshold, which we usually take to be $\epsilon_{\rm Newton}=10^{-16}$,
or until a maximum number of iterations, which we take to be 10, has been
surpassed, meaning that the Newton procedure has failed.  We use the norm
\begin{equation}
||\Usol|| \equiv \frac{1}{Ra\Delta t}
\left(||H||_\infty + \frac{1}{Pr} \max \left(||U_r||_\infty,||U_\theta||_\infty,
||U_z||\infty)\right)\right) .
\label{eq:normdef}\end{equation}

The size of the matrix representing the linear operator in \eqref{eq:BCGS} is
$(4\times 10^5)\times (4\times 10^5)$ and so the system is far too large to be
solved directly. Instead we use the BiConjugate Gradient Stabilized algorithm
\cite{Vandervorst}, which requires the right-hand-side and a procedure for
calculating the action of $\mathcal{B}_\Usol-I$ on a vector $u$.  The
right-hand-side of \eqref{eq:BCGS} is shown by \eqref{eq:difftimestep} to be
the difference between $U(t+\tdel)=B(U(t))$ and $U(t)$, i.e.~between two
(widely spaced) consecutive timesteps, while the left-hand-side is the
difference between $B_U(U(t))$ and $U(t)$, i.e.~between two {\it linearized}
timesteps.  Conjugate gradient iteration proceeds until
\begin{equation}
\frac{||(\mathcal{B}_\Usol-I)\,\usol - 
(\mathcal{B}-I)\:\Usol||}{||(\mathcal{B}-I)\:\Usol||} \leq \epsilon_{\rm BiCGS}
\end{equation}
where the threshold $\epsilon_{\rm BiCGS}$ is taken between $10^{-8}$ and
$10^{-16}$.
The reason for finding the roots of $\mathcal{B}-I$ instead of those of
$\mathcal{N}+\mathcal{L}$ is that, as shown by equation
\eqref{eq:difftimestep},
\begin{equation}
(\mathcal{B}_\Usol - I) \approx \mathcal{L}^{-1}\left(\mathcal{N}_\Usol +
  \mathcal{L}\right) \qquad\qquad \mbox{ for } \Delta t\gg 1 .
\label{eq:tdel_large}\end{equation}
This effective preconditioning by $\mathcal{L}^{-1}$ makes
$\mathcal{B}_\Usol-I$ far better conditioned than
$\mathcal{N}_\Usol+\mathcal{L}$, and greatly accelerates the convergence of
BiCGSTAB.  Note that $\Delta t\gg1$ is the limit {\it opposite} to that used
in timestepping.  We use $\tdel$ ranging between 0.2 and 10 (in contrast to
the $\tdel$ on the order of $10^{-4}$ used in temporal integration).

It is the solution of the linear system \eqref{eq:BCGS} which poses the
greatest difficulty and which determined the limits of our study.  In some
regions, convergence of BiCGSTAB required as few as 5 actions of $B_U-I$, with
more typical values ranging between 30 and 800.  In other regions, 4000
iterations did not suffice (even sometimes far from any bifurcation, where
singularity of $\mathcal{B}_\Usol-I$ is to be expected), and continuation of
the branch was eventually abandoned.

\subsection{Branch following}

In order to calculate a branch of steady states, we carry out Newton iteration
\eqref{eq:Newton} repeatedly for different values of Rayleigh
number. Generally, in the absence of turning points, one can merely use the
converged solution for one $Ra$ to initialize the Newton iteration for a
neighboring $Ra$.  This initialization procedure constitutes zero-th order
extrapolation.  We reduce the increment or decrement $\Delta Ra$ in $Ra$ if
the Newton iteration failed to converge in $N^{\rm opt}$ iterations and
increase $\Delta Ra$ if convergence took place sooner.  Specifically, if we
have computed solutions $\Usol^{(1)}$, $\Usol^{(2)}$ corresponding to
$Ra^{(1)}$, $Ra^{(2)}$ in $N^{(1)}$, $N^{(2)}$ Newton iterations, we set
\begin{eqnarray}
Ra^{(3)}&=&Ra^{(2)}+ \Delta Ra = Ra^{(2)}+\alpha
(Ra^{(2)}-Ra^{(1)}) \label{eq:ra_extrap} \\ \alpha&=&\frac{N^{\rm
    opt}+1}{N^{(2)}+1} \nonumber
\end{eqnarray}
where we take $N^{\rm opt}$ between 2 and 5.  

Linear or quadratic extrapolation in $Ra$ is easy to implement.  Assume that
converged solutions $\Usol^{(0)}$, $\Usol^{(1)}$, $\Usol^{(2)}$ have been
found for Rayleigh numbers $Ra^{(0)}$, $Ra^{(1)}$ and $Ra^{(2)}$.  We can
determine coefficients $a_i$, $b_i$, $c_i$ such that
\begin{equation}
\Usol_i = a_i Ra^2 + b_i Ra + c_i \label{eq:uquad}\\
\end{equation}
where $i$ ranges over both the gridpoints and the components
$(H,U_r,U_\theta,U_z)$.  We then use \eqref{eq:uquad} to compute an initial
condition for Newton iteration at the new value $Ra^{(3)}$ given in
\eqref{eq:ra_extrap}.  (The condition number of the $3\times 3$ system
\eqref{eq:uquad} for $a_i$, $b_i$, $c_i$ is improved if we subtract from $Ra$
the average of the three $Ra$ values.) Over many portions of many
branches we find we can easily take $\Delta Ra \geq 200$.  As an example, we
computed the marigold branch which will be described in section
\ref{sec:three-fold} from $Ra=2100$ to $Ra=18\,000$, with intervals $\Delta
Ra$ varied dynamically between 10 and 1200 according to prescription
\eqref{eq:ra_extrap}, requiring a computation time of 1200 CPU seconds on the
NEC SX-8.

Quadratic extrapolation, while not necessary for moving along a branch, proves
essential near a turning point.  Near a turning point $(Ra^{TP}, \Usol^{TP}$,
we stop using extrapolation in $Ra$, as in \eqref{eq:ra_extrap}, and instead
use extrapolation in one of the components of $\Usol$.  That is, we fix the
value of one component, $\Usol_I$, and treat $Ra$ as a dependent variable.  To
determine whether we are near a turning point, we use the fact that
\begin{equation}
\vert \Usol_i-\Usol^{TP}_i \vert\sim \sqrt{\vert Ra - Ra^{TP}\vert}
\end{equation}
so that $\Delta \Usol_i$ must eventually exceed $\Delta Ra$ as a turning point
is approached.  We monitor the relative changes by comparing the quantities
\begin{eqnarray}
\left\vert\frac{\Delta \Usol_i}{\Usol_i}\right\vert \equiv
\left\vert\frac{\Usol_i^{(1)}-\Usol_i^{(2)}}{\Usol_i^{(2)}}\right\vert
&\qquad\mbox{ with } \qquad &
\left\vert\frac{\Delta Ra}{Ra}\right\vert
\equiv\left\vert\frac{Ra^{(1)}-Ra^{(2)}}{Ra^{(2)}}\right\vert ,
\end{eqnarray}
where $\gamma$ is a multiplicative weighting factor ranging between 5 
(to favor extrapolation in $Ra$) and 0.001 (to favor extrapolation in 
$\Usol_i$). When $|\Delta \Usol_i/\Usol_i|$ exceeds $\gamma|\Delta Ra/Ra|$, 
we replace \eqref{eq:ra_extrap}, prescribing
extrapolation in $Ra$, by extrapolation in $\Usol_I$:
\begin{equation}
\Usol_I^{(3)}=\Usol_I^{(2)}+ \Delta \Usol_I =\Usol_I^{(2)}+
\alpha(\Usol_I^{(2)}-\Usol_I^{(1)}) .
\label{eq:u_extrap}\end{equation}
We use the three previous converged fields and Rayleigh numbers to determine
coefficients $a_i$, $b_i$, $c_i$ for $i\neq I$ and $a_{Ra}$, $b_{Ra}$,
$c_{Ra}$ such that
\begin{eqnarray}
\Usol_i = a_i \:\Usol_I^2 + b_i \:\Usol_I + c_i , &\qquad &
Ra = a_{Ra}\:\Usol_I^2 + b_{Ra} \:\Usol_I + c_{Ra}  \label{eq:ra_quad} 
\end{eqnarray}
and then use \eqref{eq:ra_quad} to compute a new $Ra$ and $\Usol_i,i\neq I$
corresponding to the $\Usol_I$ prescribed by \eqref{eq:u_extrap}.  This allows
us to change direction in $Ra$; \eqref{eq:ra_quad} may lead to
$Ra^{(3)}-Ra^{(2)}$ of opposite sign to that of $Ra^{(2)}-Ra^{(1)}$, unlike in
equation \eqref{eq:ra_extrap}.

The procedure above treats $\Usol_I$ as an independent variable (as prescribed
in equation \eqref{eq:u_extrap}) and $Ra$ as a dependent variable (as
prescribed in equation \eqref{eq:ra_quad}) in the predictor step
(initialization).  In this investigation, we have left the corrector step
(Newton iteration) unchanged, that is, $Ra$ remains unaltered by
\eqref{eq:BCGS}.  One strategy we have employed is to relax the tolerances
near the turning point, for example to $\epsilon_{\rm Newton}=10^{-13}$ and
$\epsilon_{\textsc BCGS}=10^{-8}$.  Like Xin~\cite{Xin}, we have succeeded in
traversing a number of turning points in this way, despite 
the near-singularity of the matrix $(B_\Usol-I)$ near a bifurcation point.

\subsection{Linear stability analysis}
\label{sec:linear_stability}

Once branches have been computed, we wish to determine their stability.
In order to perform linear stability analysis of a steady state
$\Usol\equiv(H,\bU)$, we carry out
time integration of the Boussinesq equations linearized about 
$\Usol$ for an infinitesimal perturbation $\usol\equiv(h,\bu)$:
\begin{equation}
\frac{d\usol}{dt} = \mathcal{L}\usol + \mathcal{\mathcal{N}_\Usol}\usol
\label{eq:linschematic}\end{equation}
We use the same timestepping formula \eqref{eq:timestep} as for the nonlinear
problem:
\begin{equation}
\usol(t+\Delta t) = (I-\Delta t \mathcal{L})^{-1}(I + \Delta t
\mathcal{N}_\Usol) \usol(t) \equiv \mathcal{B}_\Usol\usol(t)
\label{eq:lintimestep}\end{equation}
by carrying out the substitutions in \eqref{eq:jacobian}. Since
\begin{equation}
\mathcal{B}_\Usol \approx \exp(\Delta t (\mathcal{L}+\mathcal{N}_\Usol))
\qquad\qquad \mbox{ for } \Delta t\ll 1 ,
\label{eq:expon}
\end{equation}
the eigenvalues $\lambda$ of $\mathcal{B}_\Usol$ and eigenvalues $\sigma$ of
$\mathcal{L}+\mathcal{N}_\Usol$ are related via
\begin{equation}
\lambda\approx \exp(\sigma\Delta t) \Longleftrightarrow\sigma \approx \frac{1}{\Delta t} \log |\lambda| \qquad\qquad \mbox{ for } \Delta t\ll 1
\end{equation}
The stability of $\,\Usol$ is determined by the sign of the leading eigenvalue
$\sigma_{\max}$ (that with largest real part) of
$\mathcal{L}+\mathcal{N}_\Usol$, which corresponds to the dominant eigenvalue
$\lambda_{\max}$ (that with largest magnitude) of $\mathcal{B}_\Usol$.  Equation
\eqref{eq:lintimestep} prescribes acting with the linear operator
$\mathcal{B}_\Usol$ on $\usol(t)$; when repeated over many $\Delta t$'s,
$\usol$ will converge to the eigenvector corresponding to $\lambda_{\max}$,
which is itself approximated by the Rayleigh quotient
\begin{equation}
\lambda_{\max} \approx \lim_{t\rightarrow\infty}\frac{\langle \usol(t),
  B_\Usol \usol(t) \rangle}{\langle \usol(t), \usol(t) \rangle} .
\end{equation}

To determine several leading eigenvalues, the power method is generalized to
the Arnoldi-Krylov method~\cite{Arnoldi}.  This consists of orthonormalizing a
small number of fields
\begin{equation}\{ \usol(t=0),\usol(t=T),\usol(t=2T)\dots \usol(t=(K-1)T)\}
\label{eq:Kspace}\end{equation}
to create vectors $v_1, v_2, v_3, \ldots v_K$, and then a small Hessenberg
matrix $H_{jk}\equiv \langle v_j,\mathcal{B}_\Usol v_k\rangle$, which is
directly diagonalised.  Its eigenvalues approximate eigenvalues $\lambda$ of
$\mathcal{B}_\Usol$, while its eigenvectors $\phi$ consist of coefficients of
the vectors $v_j$, to be combined to form approximate 
eigenvectors $\Phi\equiv \sum_j \phi_j v_j$ of $\mathcal{B}_\Usol$.
The accuracy of these approximate eigenpairs is measured by the residue
$||\mathcal{B}_\Usol\Phi - \lambda \Phi||$ 
in the case of complex eigenvalues.  
The integration of \eqref{eq:lintimestep}
is continued until the residues of the desired eigenvalues are below some
acceptance criterion, usually near $10^{-6}$.

The timestep required is similar to that for timestepping.  One obvious
restriction comes from the explicit scheme used in \eqref{eq:lintimestep} for
$\mathcal{N}_\Usol$; a timestep which violates this stability requirement
leads to approximate eigenvalues of $\mathcal{B}_\Usol$ which bear no
resemblance to those of $\exp(\tdel(\mathcal{L}+\mathcal{N}_\Usol))$.  For
smaller $\tdel$, the accuracy of the eigenvalues depends on $\tdel$ because of
the approximation \eqref{eq:expon}.  In particular, the time-splitting error
means that $\mathcal{B}_\Usol$ is not a function of
$\mathcal{L}+\mathcal{N}_\Usol$.
(In contrast, the errors in the pitchfork and turning point bifurcation 
thresholds obtained by Newton's method result only from the spatial 
discretization.)
We have used $\tdel=10^{-3}$ for $Ra\lesssim 10\,000$ and $\tdel=5\times
10^{-4}$ for $Ra>sim 10\,000$.  
We estimate our accuracy in locating bifurcation points and 
stability ranges to be $\Delta Ra \lesssim 1$.
We used $K=10$ vectors and a
time interval of $T=100\Delta t$, i.e.~$T=0.1$ or $T=0.05$ to create the
Krylov vectors \eqref{eq:Kspace}, and an acceptance criterion for the residues of
$10^{-6}$.

A method which produces approximate eigenvalues which are independent of
$\tdel$ is the inverse Arnoldi method~\cite{Timesteppers}, in which
\eqref{eq:lintimestep} is replaced by
\begin{equation}
\usol^{n+1} = (\mathcal{L}+\mathcal{N}_\Usol)^{-1}\usol^n.\label{eq:inverse_arnoldi_1}
\end{equation}
This is accomplished in practice by solving the equation 
\begin{equation}
(\mathcal{B}_\Usol-I)\usol^{n+1}=\tdel(I-\tdel \mathcal{L})^{-1} \usol^n \label{eq:inverse_arnoldi_2}
\end{equation}
The equivalence between \eqref{eq:inverse_arnoldi_1} and
\eqref{eq:inverse_arnoldi_2} follows from a calculation similar to
\eqref{eq:difftimestep}.  Equation \eqref{eq:inverse_arnoldi_2} is very
similar to \eqref{eq:BCGS} and is also solved by BiCGSTAB. Only a few
iterations (between 1 and 10) of \eqref{eq:inverse_arnoldi_2} lead to an
extremely accurate eigenvalue.  However, the inverse Arnoldi is more difficult
to implement than the exponential Arnoldi method.
For this reason, we have chosen not to do so for this study.

\section{Results}
\label{sec:results}

\subsection{Bifurcation Diagram}

\begin{figure}[htp]
\vspace*{-1.3cm}
\begin{center}
\includegraphics[width=17cm]{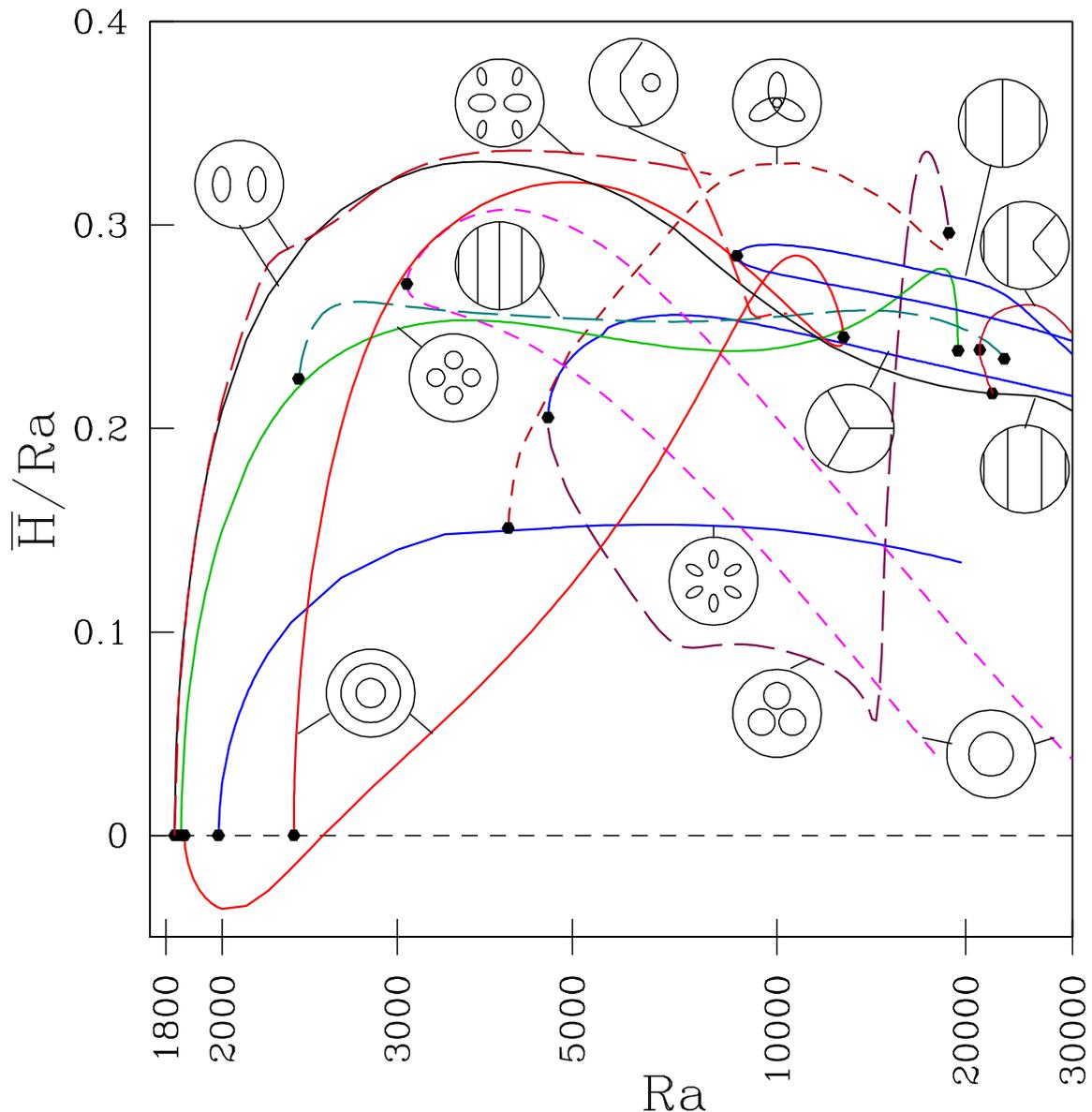}
\end{center}
\vspace*{-0.5cm}
\caption{(Color online) Bifurcation diagram containing 17 branches of steady states, in
  addition to the conductive branch (indicated by short-dashed horizontal
  line).  Shown are pizza (solid green), four-roll (long-dashed turquoise),
  two-tori (solid red; 2), torus (long-dashed magenta; 2), marigold (solid
  blue), mitsubishi (short-dashed purple), cloverleaf (long-dashed purple) and
  mercedes (solid blue), three-roll (solid black), tiger (long-dashed brick),
  asymmetric three-roll (solid brick; 2), two-roll (solid blue; 2), CO
  (long-dashed red) branches.
The notation ``torus (long-dashed magenta; 2)'', e.g., signifies
that there are 2 torus branches, related by saddle-node bifurcations and both
shown as long-dashed magenta curves.
Note that the bifurcation diagram gives no information concerning stability;
i.e.~whether a solution curve is depicted as solid or dashed does not indicate
its stability.
  Turning points or pitchfork bifurcations are shown as dots.}
\label{fig:bifdiag_full}
\end{figure}

Using the methods described in section \ref{sec:numerical}, we have succeeded
in continuing the branches we found previously via time 
integration~\cite{Boronska_PhD,Boronska_PRE1}.  
By going around turning points and bifurcation points,
we have computed a total of 17 branches of convective steady states.  These
are related to the conductive state and to each other by 5 primary and 3
secondary pitchfork bifurcations, and 8 saddle-node bifurcations.  The
bifurcation diagram is shown in figure \ref{fig:bifdiag_full}.  Tables
summarizing all of the branches and bifurcations we have found are given in
section \ref{sec:conclusion}.

The axes of figure \ref{fig:bifdiag_full} have been chosen with care.  In
order to show the full extent of our calculations in Rayleigh number and, at
the same time, distinguish between various low-Rayleigh-number primary
bifurcations, figure \ref{fig:bifdiag_full} uses a logarithmic scale in $Ra$.
More specifically, using $\log(Ra-1000)$ distinguishes primary bifurcations 
better than $\log Ra$.  The vertical axis was
chosen to best distinguish between the various branches.  The quantity $\Hmax$
is the maximum absolute value of the temperature deviation over the ring at
$(r=0.3,\theta,z=0)$
\begin{equation}
\Hmax \equiv \max_\theta | H(r=0.3,\theta,z=0) | .
\label{eq:Hdef}\end{equation}
$\Hmax$ itself and the commonly used Nusselt number deviation
\begin{equation}
Nu -1 =\int r dr \: d\theta \: \partial_z H (r,\theta,z=0) \: -1
\end{equation}
have a strong linear dependence on $Ra$; plotting them directly as a function
of $Ra$ does little to separate the branches.  We have therefore chosen
instead to represent each state by its value of $\Hmax/Ra$.  (Exceptionally,
for the first two-tori branch, we have plotted $-\Hmax/Ra$ for low $Ra$ to
avoid a reversal in slope due to the absolute value
in \eqref{eq:Hdef}.)  Each branch in figure \ref{fig:bifdiag_full} is a
representative of a number of branches -- the group orbit -- that can be
obtained by reflection and rotations.

In order to understand the complicated bifurcation diagram in 
figure \ref{fig:bifdiag_full}, we will select various aspects for
detailed study below.

\subsection{Primary Bifurcations}

We give in table \ref{tab:firstbif} the first critical wavenumber and
Rayleigh number pairs.
The thresholds given to $\Delta Ra=0.1$ are extrapolations from the branches
calculated using Newton's method.  The thresholds given as integer values were
calculated from the linear stability analysis of the
conductive branch. In the remainder of the manuscript, we round Rayleigh
numbers to integer values (except in a few very specific cases).  Our
thresholds agree quite well with those of previous researchers.  The
discrepancies are typically on the order of $0.3\%$ with the calculations of
Ma \etal~\cite{Ma} and on the order of $0.02\%$ with those of
Martin-Witkowski~\cite{Laurent}, which we believe to be the two most recent
threshold calculations in this geometry.  With increasing Rayleigh number,
many other bifurcations occur from the conductive state, both to higher
wavenumbers and to different eigenmodes with the same wavenumbers.  The
branches created at such bifurcations are necessarily unstable.

\begin{table}[htp]
\begin{center}
\begin{tabular}{|c||c|c|c|c||c|c|c|c|c|}
\hline
$Ra$ ~&~~ 1828.4 ~~&~~ 1849.4 ~~&~~ 1861.6 ~~&~~ 1985.3
&~~ 2055 ~~&~~ 2172 ~~& ~~2255~~ & ~~2328.0 \\\hline
$m$ & 1 & 2 & 0 & 3 & 4 & 5 & 1 & 0  \\\hline
\end{tabular}
\end{center}
\caption{(Color online) First bifurcations from conductive state}
\label{tab:firstbif}
\end{table}
\begin{figure}[!htp]
\begin{center}
\includegraphics[width=12cm]{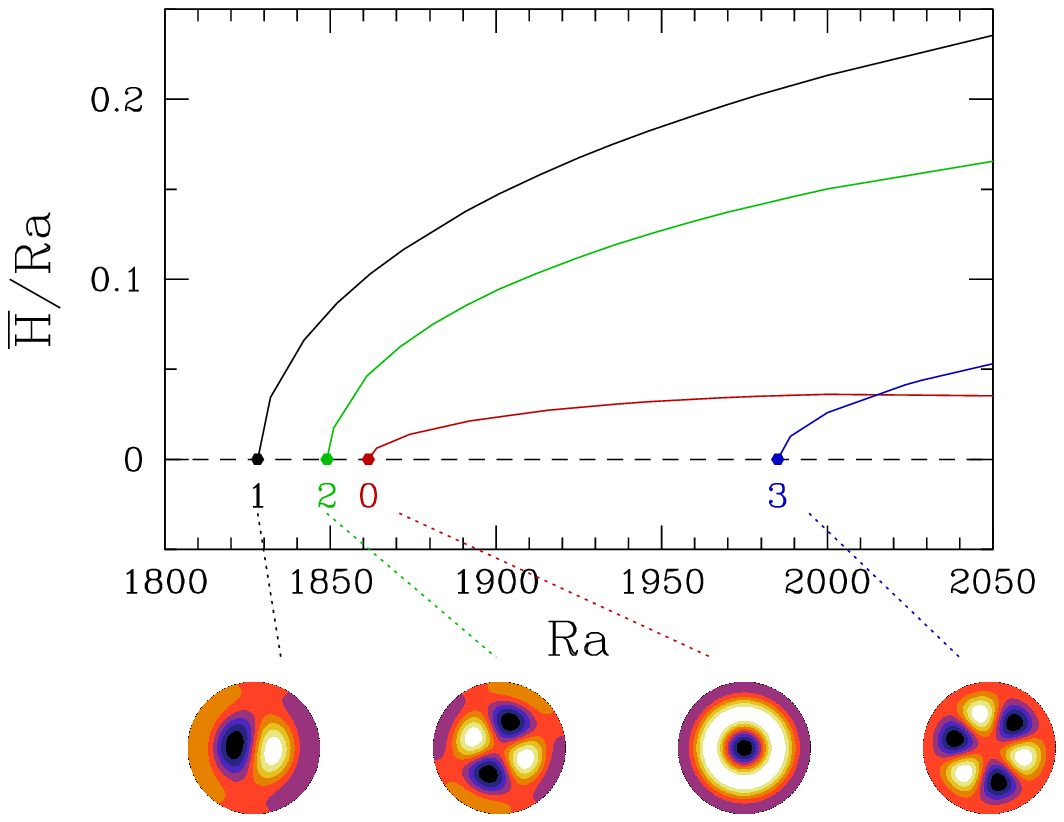}
\end{center}
\caption{(Color online) Primary branches bifurcating from conductive state.
For this aspect ratio, $\Gamma=2$, and with insulating lateral walls,
the first four critical wavenumber and Rayleigh number pairs are
($m=1$, $Ra=1828$; black), ($m=2$, $Ra=1849$; green), ($m=0$, $Ra=1861$; red), 
and ($m=3$, $Ra=1985$; blue). Below are representative states from each of the
bifurcating branches.}
\label{fig:bifcond}
\end{figure}

It is the first four bifurcations of table \ref{tab:firstbif}, along with the
last column, which will prove relevant to the steady states observed,
i.e.~the stable ones.  In figure \ref{fig:bifcond} we show these
first bifurcations, along with corresponding nonlinear states at slightly
supercritical values of $Ra$. We recognize the dipole, pizza and two-tori
states.  The other states in figures \ref{fig:dns:neu} and
\ref{fig:bifdiag_full} -- the two-, three-, and four-roll states, or the
torus, mercedes, and CO states -- are not present in figure
\ref{fig:bifcond}. Their origin is addressed in the sections which follow.

\subsection{Pizza and Four-roll branches ($\mathbf{m=2}$)}

\begin{figure}
\vspace*{-1cm}
\includegraphics[width=12cm]{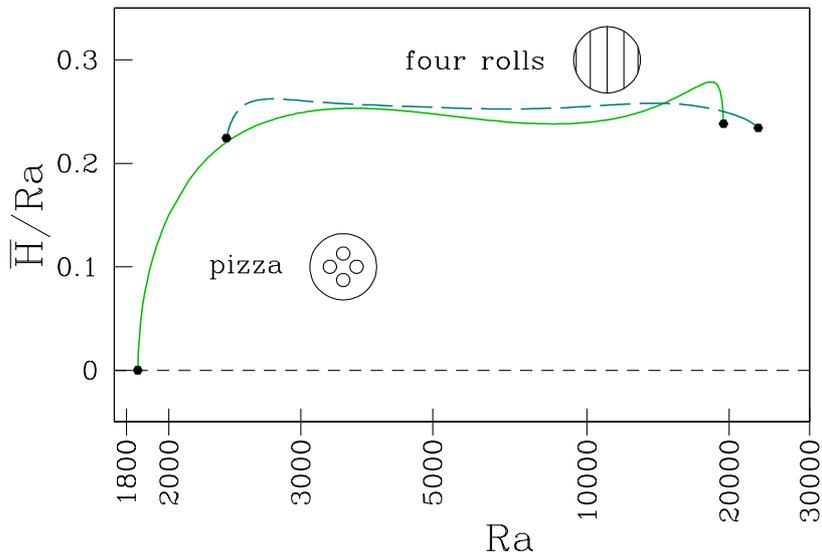}
\caption{(Color online) Partial bifurcation diagram including $m=2$ primary branch and
  connecting branches. Bifurcations are shown as dots.  The primary {\bf
    Pizza} branch (solid, green) bifurcates from the conductive state at
  $Ra=1849$ and terminates in a saddle-node bifurcation at $Ra\approx
  19\,450$. The {\bf Four-roll} branch (dashed, turquoise) bifurcates from the
  pizza branch at $Ra=2353$ and terminates at a saddle-node bifurcation at
  $Ra\approx 23\,130$.}
\label{fig:pizza_sm}
\end{figure}

\begin{figure}
\includegraphics[width=17cm]{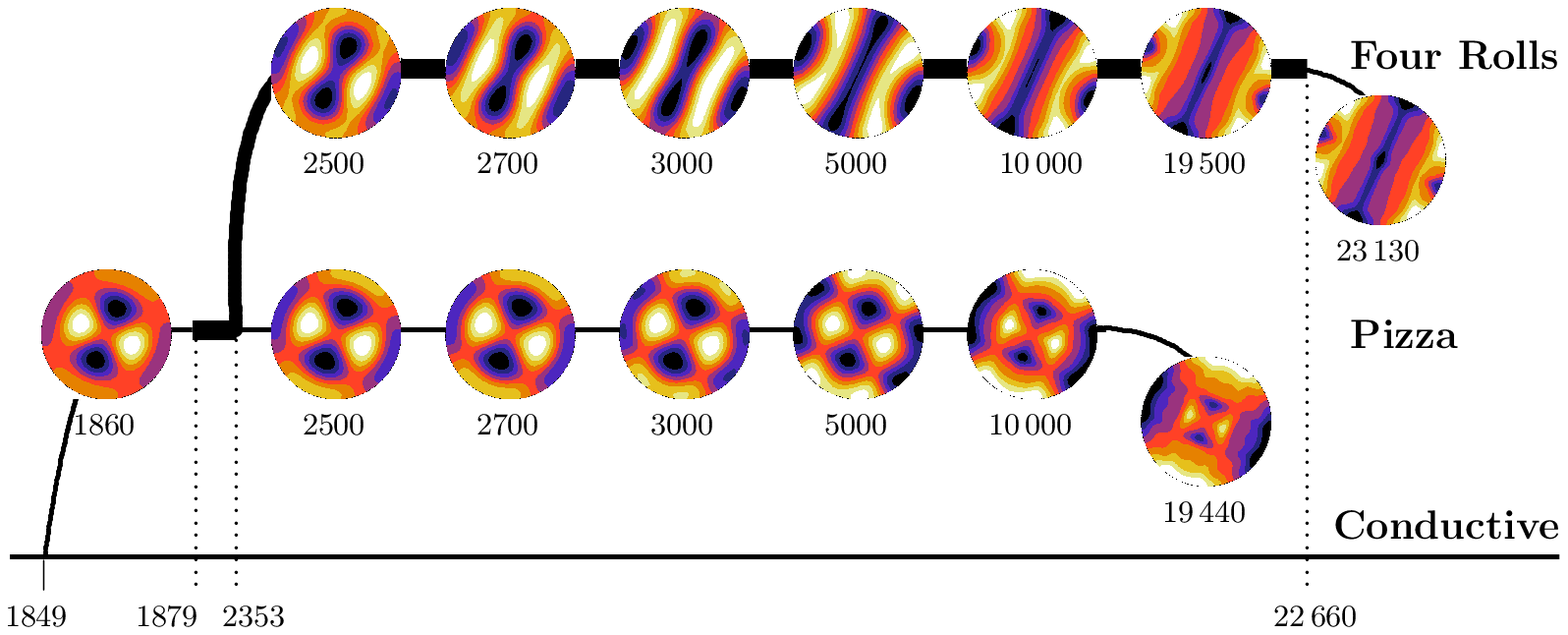}
\caption{(Color online) Schematic partial bifurcation diagram relating branches originating
  from the $m=2$ bifurcation.  At $Ra=1849$ the {\bf Pizza} branch originates
  via a circle pitchfork bifurcation from the conductive state corresponding
  to an $m=2$ eigenvector. It terminates at a turning point at $Ra\lesssim
  19\,450$ and is stable for $1879 \leq Ra \leq 2353$.  At $Ra = 2353$, a
  secondary pitchfork bifurcation leads to a {\bf Four-roll} branch, which is
  stable for $Ra\lesssim 22\,660$ and ends at a turning point at $Ra\approx
  23\,130$.  For visual clarity, the Rayleigh numbers given for the
  representative states have been rounded to the nearest 10, 100 or 1000.}
\label{fig:pizza_pic}
\end{figure}

\begin{figure}
\includegraphics[width=17cm]{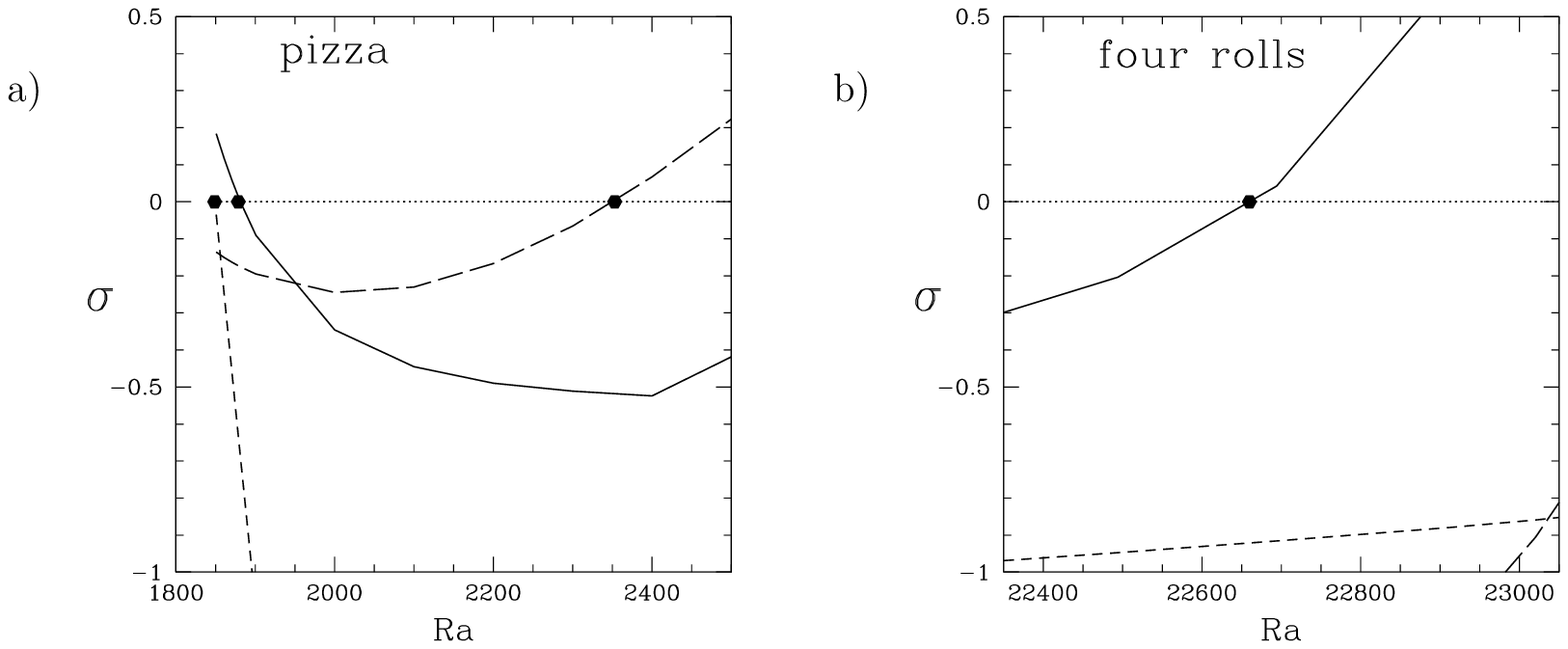}
\caption{Four leading eigenvalues of (a) the pizza branch at low $Ra$ and of
  (b) the four-roll branch at high $Ra$. Bifurcations (zero crossings)
  indicated by dots.  The zero eigenvalue (dotted) which exists
  throughout corresponds to the marginal stability to rotation of the pattern.
  a) The bifurcating eigenvalue (short-dashed) decreases steeply from 0
  at onset, $Ra=1849$. The pizza branch is initially unstable since it
  inherits the unstable eigenvalue (solid) of the conducting branch,
  due to the preceding $m=1$ bifurcation.  This leading eigenvalue decreases
  with $Ra$, crossing zero at $Ra=1879$. Another eigenvalue (long-dashed)
  becomes positive at $Ra=2353$, accompanying the bifurcation to the four-roll
  branch.  The stability interval of the pizza branch is $1879 \leq Ra\leq
  2353$.  b) The four-roll branch loses stability near $Ra=22\,660$.}
\label{fig:pizza_eigs}
\end{figure}

We now focus on various sets of solution branches.  We begin with the branches
arising from the instability to an $m=2$ quadrupolar eigenvector at $Ra=1849$,
because these are free from the complications which we will encounter for the
other azimuthal wavenumbers.  We use three figures to describe the structure
of these branches.  Figure \ref{fig:pizza_sm} uses the same coordinates as
figure \ref{fig:bifdiag_full}, merely extracting the relevant branches.
Figure \ref{fig:pizza_pic} is a qualitative bifurcation diagram, accompanied
by illustrations of representative states along the branches.  Finally, figure
\ref{fig:pizza_eigs} shows leading eigenvalues, from which the stability of
the underlying branches can be deduced.

The bifurcation sequence is best understood by studying figure
\ref{fig:pizza_pic}. The schematic quantity along the vertical axis and the
monotonic but non-uniform Rayleigh-number progression along the horizontal
axis, are chosen to separate the different branches and to illustrate the
bifurcations.  Representative states along the branches are illustrated via
temperature distributions in the midplane ($z=0$), with light portions
representing hot rising fluid and dark portions representing cold desceding
fluid.  To avoid further cluttering the figure, the Rayleigh numbers given for
the representative states have been rounded to the nearest 10, 100 or even
1000, with precise bifurcation points given along the axis.  The azimuthal
orientation of the representative states is arbitrary, and to each branch
corresponds another branch obtained by the Boussinesq reflection symmetry
which, for these illustrations, would mean reversing light and dark.

A circle pitchfork bifurcation from the {\bf Conductive} branch to an $m=2$
eigenmode takes place at $Ra=1849$.  Figure \ref{fig:pizza_pic} shows that,
near onset, the states along the branch created by this bifurcation contain
two hot upwelling spots and two cold downwelling spots.  Their resemblance to
a small pizza leads us to call this the {\bf Pizza} branch.  As $Ra$
increases, the central convective regions shrink.  By the time the pizza
branch terminates at a saddle-node bifurcation at $Ra=19\,450$, most of the
convection takes place at four regions along the edge of the container.

A pitchfork bifurcation at $Ra=2353$ from the pizza branch breaks the symmetry
between hot upwelling and cold downwelling fluid: the two downwelling spots
merge as the two upwelling spots elongate (or vice versa for the complementary
branch, not shown).  This secondary bifurcation is also computed by Ma
\cite{Ma}, who gave its threshold as $Ra=2350$.  The pitchfork bifurcation
leads to a {\bf Four-roll} branch which terminates at a saddle-node
bifurcation at $Ra\approx 23\,130$.  Along the four-roll branch, the
convective regions diminish as $Ra$ increases, as was the case for the pizza
branch; the rolls become wide, with narrow upwelling and downwelling
boundaries.

Figure \ref{fig:pizza_eigs}\textit{a} shows the four leading eigenvalues of
the pizza branch near onset, computed by the methods described in section
\ref{sec:linear_stability}. They are grouped into distinct sets by examining
the spatial structure, especially the azimuthal wavenumber spectrum, of the
corresponding eigenvectors, and then plotted as curves.  Very near onset, each
eigenvalue can be associated with an azimuthal wavenumber, since it is
connected to an eigenvalue of the conductive branch.  The zero eigenvalue
($m=2$, sometimes called the phase mode) which exists throughout corresponds
to the marginal stability to rotation of the pattern.  The eigenvalue which is
zero at onset and then rapidly decreases is that corresponding to the circle
pitchfork which creates this branch (also $m=2$, sometimes called the
amplitude mode).  The positive eigenvalue ($m=1$) at onset results from the
fact that the bifurcation to the dipole branch at $Ra=1828$ precedes the
creation of the pizza branch. The pizza branch inherits this instability when
it is created at $Ra=1849$ and becomes stable at $Ra=1879$, when the leading
eigenvalue becomes negative, as shown in figure
\ref{fig:pizza_eigs}\textit{a}.  This confirms our time-dependent
simulations~\cite{Boronska_PhD,Boronska_PRE1}, summarized in figure
\ref{fig:dns:neu}, which shows the pizza branch as stable at $Ra=2000$.
However, it enjoys only a short $Ra$-interval of stability, as another
eigenvalue (connected to the $m=0$ eigenvalue of the conductive branch)
becomes positive at $Ra=2353$, when the secondary pitchfork bifurcation
creates the four-roll branch.

Figure \ref{fig:pizza_eigs}\textit{b} shows that the four-roll branch remains
stable until $Ra\lesssim 22\,660$, a far wider Rayleigh-number interval than
the pizza branch.  Indeed, roll states are preferred by convective systems and
are those generally observed in experiments and time-dependent
simulations.  In particular, a four-roll state was computed in time-dependent
simulations~\cite{Boronska_PhD,Boronska_PRE1} for $Ra$ between 5000 and
$20\,000$ (see figure \ref{fig:dns:neu}) and is one of the five states
observed experimentally by Hof \etal~\cite{Hof} at $Ra=14\,200$.

As explained in section \ref{sec:symmetries}, the four-roll states have
symmetry group $D_2$: they are invariant under rotation by $\pi$ and
reflection in either of the symmetry axes, as stated in
\eqref{eq:dmrot}--\eqref{eq:dmref}.  The pizza states are also invariant under
the additional symmetry given in \eqref{eq:dmz} (rotation by $\pi/2$,
$z\rightarrow -z$, $U_z\rightarrow -U_z$, $H\rightarrow -H$) so have the
larger symmetry group $D_2\times Z_2$, as is typical for the primary branches
bifurcating from the conductive state.

\subsection{Torus and two-tori branches ($\mathbf{m=0}$)}

\begin{figure}
\includegraphics[width=12cm]{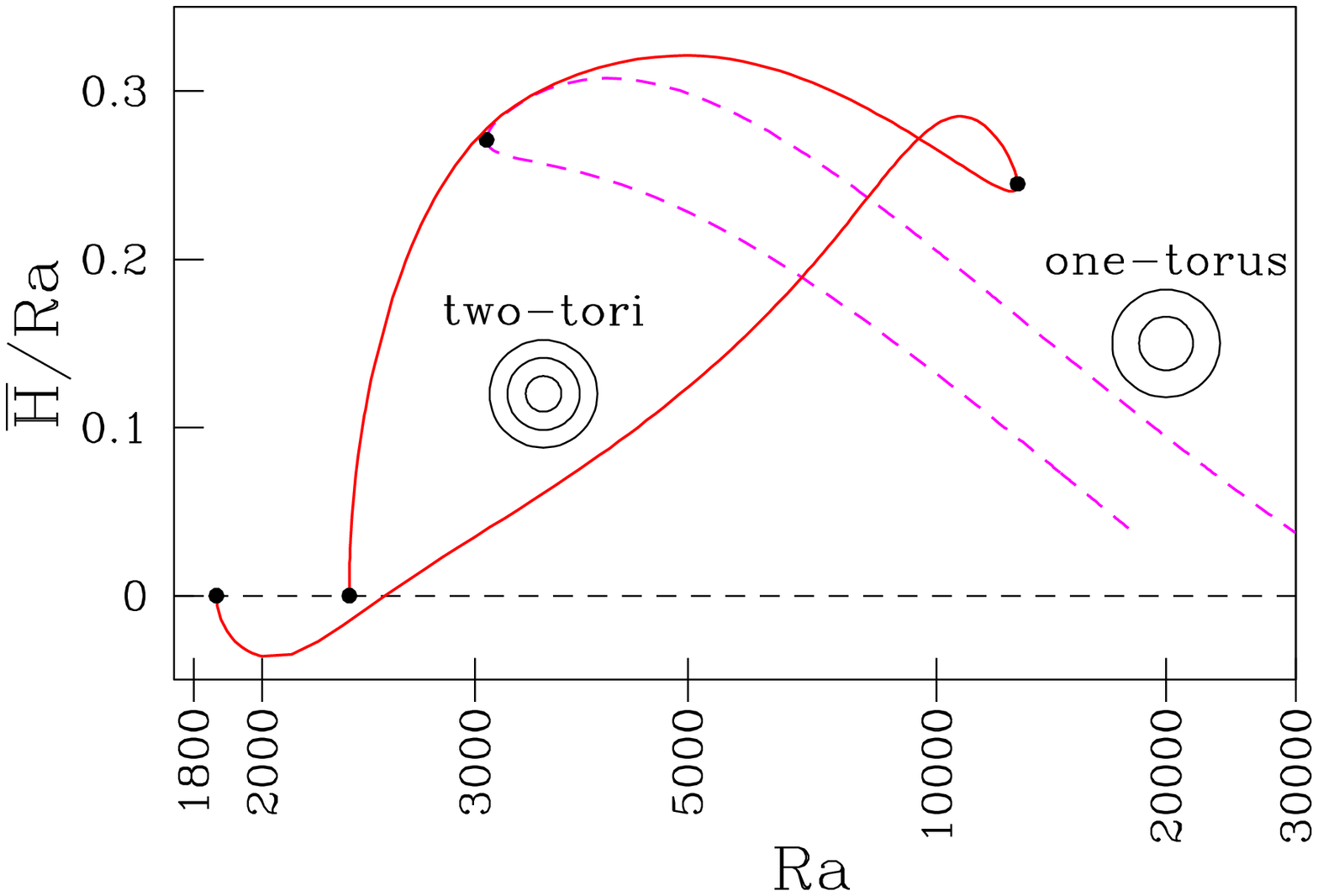}
\caption{(Color online) Partial bifurcation diagram including only axisymmetric states.
The {\bf Two-tori} branches (dashed, magenta) emerge from pitchfork 
bifurcations from the conductive branch at
$Ra=1861.5$ and $Ra=2328$. They join and terminate at a turning point at
$Ra=12\,711$. The {\bf One-torus} branches (solid, red) emerge 
from a turning point at
$Ra=3076$ and seem to be unconnected to the conductive state.}
\label{fig:axi_sm}
\vspace*{-1cm}
\includegraphics[width=17cm]{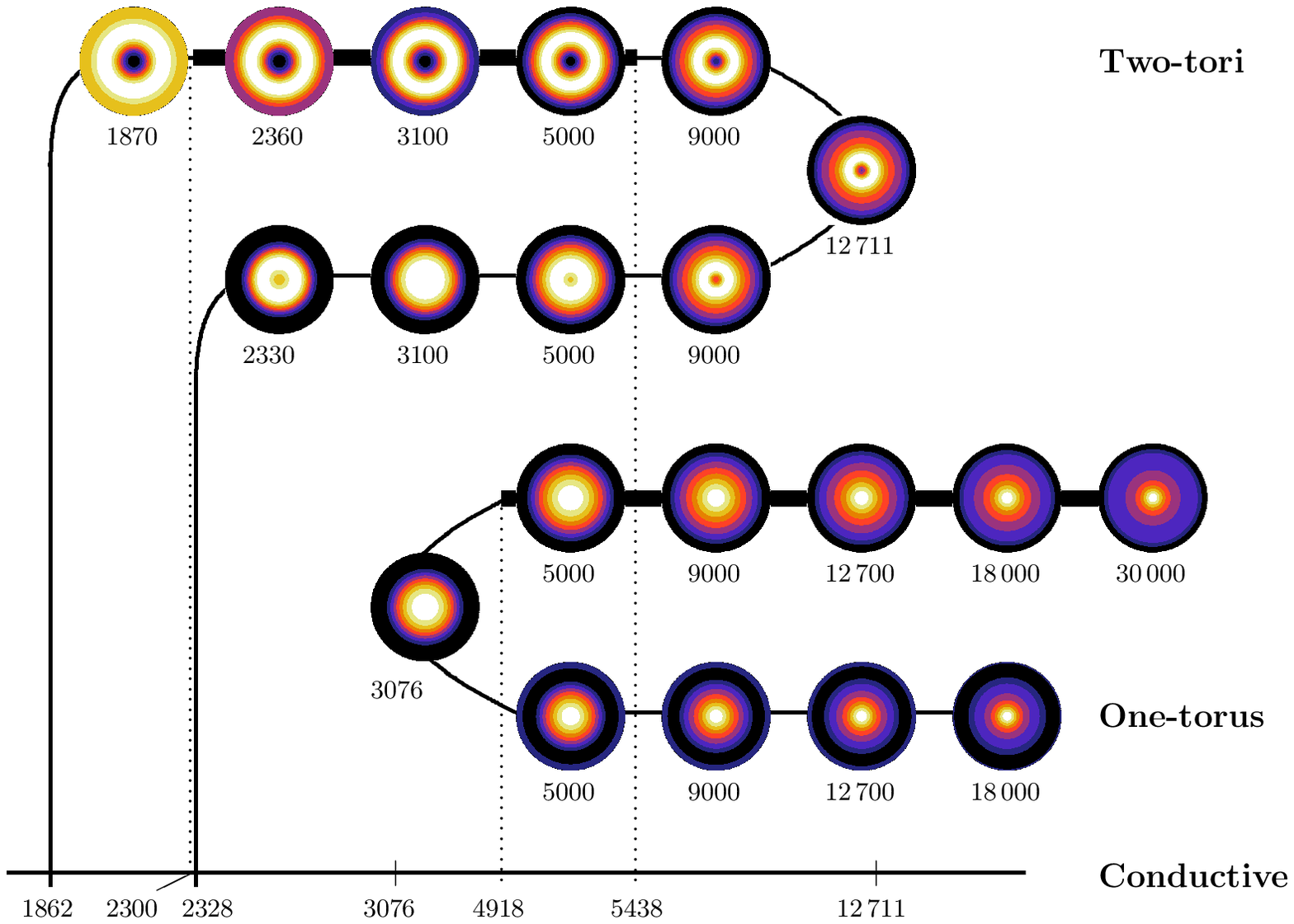}
\caption{(Color online) Schematic partial bifurcation diagram showing axisymmetric branches.
  The {\bf Two-tori} branches are connected to the conductive branch via
  pitchfork bifurcations at $Ra=1862$ and 2328, and to each other via a
  turning point at $Ra=12\,711$.  The upper two-tori branch is stable for
  $2300\leq Ra \leq 5438$.  The {\bf One-torus} branches are connected to each
  other via a turning point at $Ra=3076$. The upper one-torus branch is stable
  for $Ra\geq 4918$.  Most states along the two-tori branches contain two
  concentric rolls, but states on the lower two-tori branch resemble those on
  the upper one-torus branch for $Re\leq 3500$.}
\label{fig:axi_pic}
\end{figure}

\begin{figure}
\includegraphics[width=17cm]{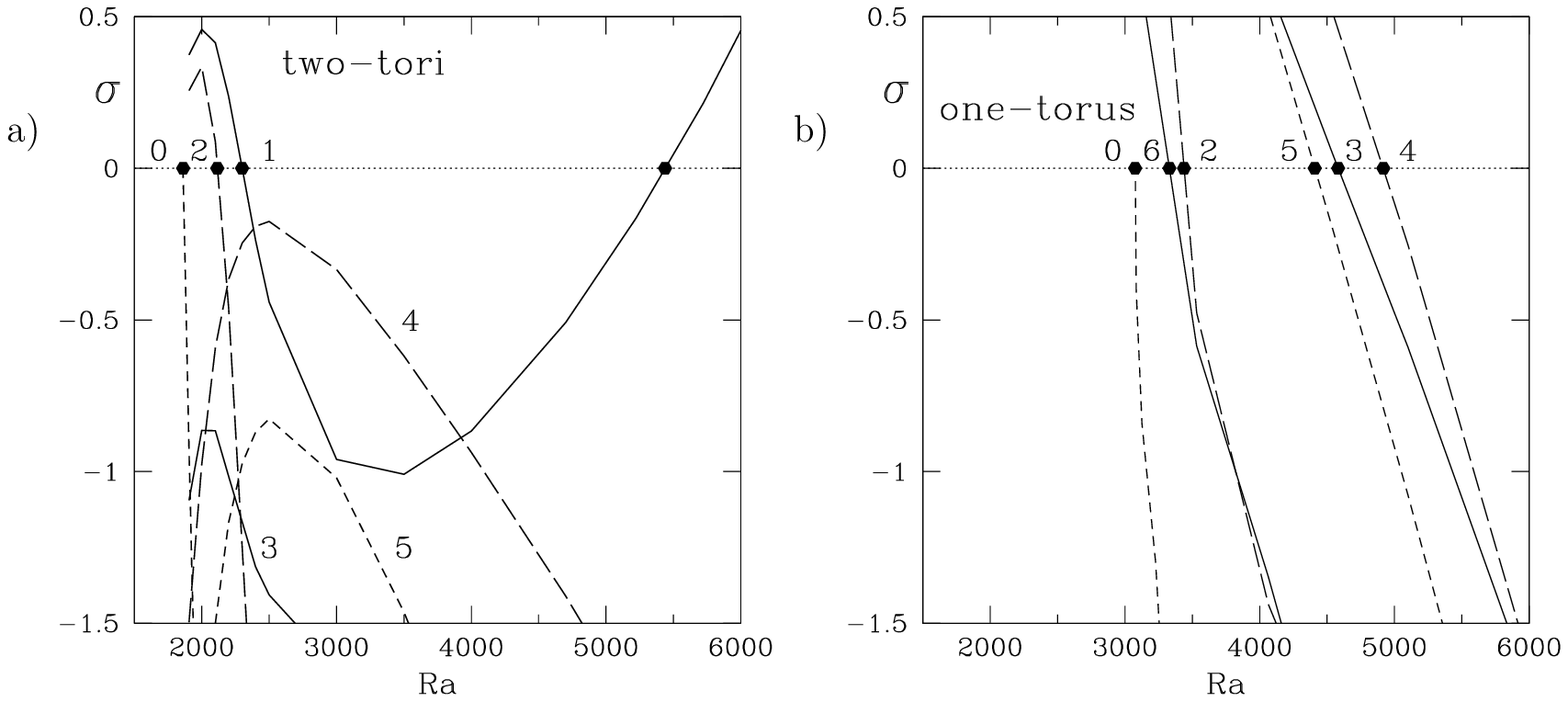}
\caption{Five leading eigenvalues of upper axisymmetric branches, 
each labelled with its azimuthal wavenumber $m$.
  a) The upper
  two-tori branch has two positive eigenvalues at onset at $Ra=1862$, which
  cross zero at $Ra=2116$ and $Ra=2300$ and again at $Ra=5438$.  It is stable
  for $2300\leq Ra \leq 5438$.  b) The upper one-torus branch has five
  positive eigenvalues at onset at $Ra=3076$.  These cross zero at $Ra=$3330,
  3438, 4408, 4582 and 4918, above which the branch is stable.}
\label{fig:axi_eigs}
\end{figure}

We now survey the axisymmetric branches.  Figure \ref{fig:axi_sm} extracts the
axisymmetric branches from the complete bifurcation diagram of figure
\ref{fig:bifdiag_full}.  There are two pairs of branches, i.e.~a total of four
branches of axisymmetric states.  A schematic bifurcation diagram showing
representative states is given in figure \ref{fig:axi_pic} and leading
eigenvalues are shown in figure \ref{fig:axi_eigs}.


The {\bf Two-tori} branches result from two pitchfork bifurcations from the
conductive state at $Ra=1862$ and $Ra=2328$.  These two branches meet and
annihilate at a turning point at $Ra=12\,711$.  Most of the states along these
branches contain two concentric toroidal convection rolls.  The branch created
at $Ra=1862$, which we call the upper or stable two-tori branch, is the more
stable of the two.  In fact, it is unstable when it is first created, as shown
in figure \ref{fig:axi_eigs}, since the $m=1$ and $m=2$ bifurcations precede
the $m=0$ bifurcation.  For an axisymmetric convective state, the eigenvectors
are each associated with a single azimuthal wavenumber $m$. 
The bifurcating eigenvalue, with $m=0$, is 0 at onset and rapidly decreases.
One of the two
leading eigenvalues becomes negative at $Ra=2116$ ($m=2$) and the second at
$Ra=2300$ ($m=1$), stabilizing this two-tori branch.  These stabilizing
bifurcations were also computed by Ma \etal~\cite{Ma}, with thresholds 2113
and 2245, respectively.
The upper two-tori branch remains stable until $Ra=5438$, when the $m=1$
eigenvalue becomes positive again.

The {\bf One-torus} branches emerge from a saddle-node bifurcation at
$Ra=3076$. The states on these branches all contain a single toroidal
convection roll.  We have not found any connection between these branches and
any others, including the conductive branch.  Both branches are initially
unstable.  The lower one-torus branch never stabilizes and we have been unable
to calculate it past $Ra=17\,857$.  The upper or stable one-torus branch is
created with five positive eigenvalues.  As $Ra$ increases, these successively
become negative, as shown in figure \ref{fig:axi_eigs}. The branch is stable
for $Ra\geq 4918$ and exists until at least $Ra=29\,940$.  It is clear that
the axisymmetric state observed at $Ra=14 200$ in experiment~\cite{Hof} and in
time-dependent simulation~\cite{Boronska_PhD,Boronska_PRE1,Leong,Ma} must be
on the stable one-torus branch, and not on the two-tori branch (which is
unstable for $Ra>5438$ and does not exist for $Ra>12\,711$) which bifurcates
from the conductive state.

Along the upper two-tori branch the inner roll dominates, while along the
lower two-tori branch, the outer roll dominates.  These states do not
necessarily all contain two rolls.  In particular, some states along the lower
two-tori branch for $Ra\lesssim 5000$ seem to contain only one roll.  These
states bear a qualitative and quantitative resemblance to those on the upper
one-torus branch. On figure \ref{fig:axi_sm}, the upper one-torus and the
lower two-tori branches are nearly tangent to one another over the interval
$3000 \lesssim Ra \lesssim 3500$, while at the turning point at $Ra=3076$, the
one-torus states bear a strong resemblance to the lower two-tori branch.

The axisymmetric convective branches break the Boussinesq symmetry
\eqref{eq:ztempsym}, while retaining the $O(2)$ azimuthal symmetry
\eqref{eq:o2rot}-\eqref{eq:o2ref}.

\subsection{Mercedes, Cloverleaf, Mitsubishi and Marigold states ($\mathbf{m=3}$)}
\label{sec:three-fold}

\begin{figure}[htp]
\includegraphics[width=12cm]{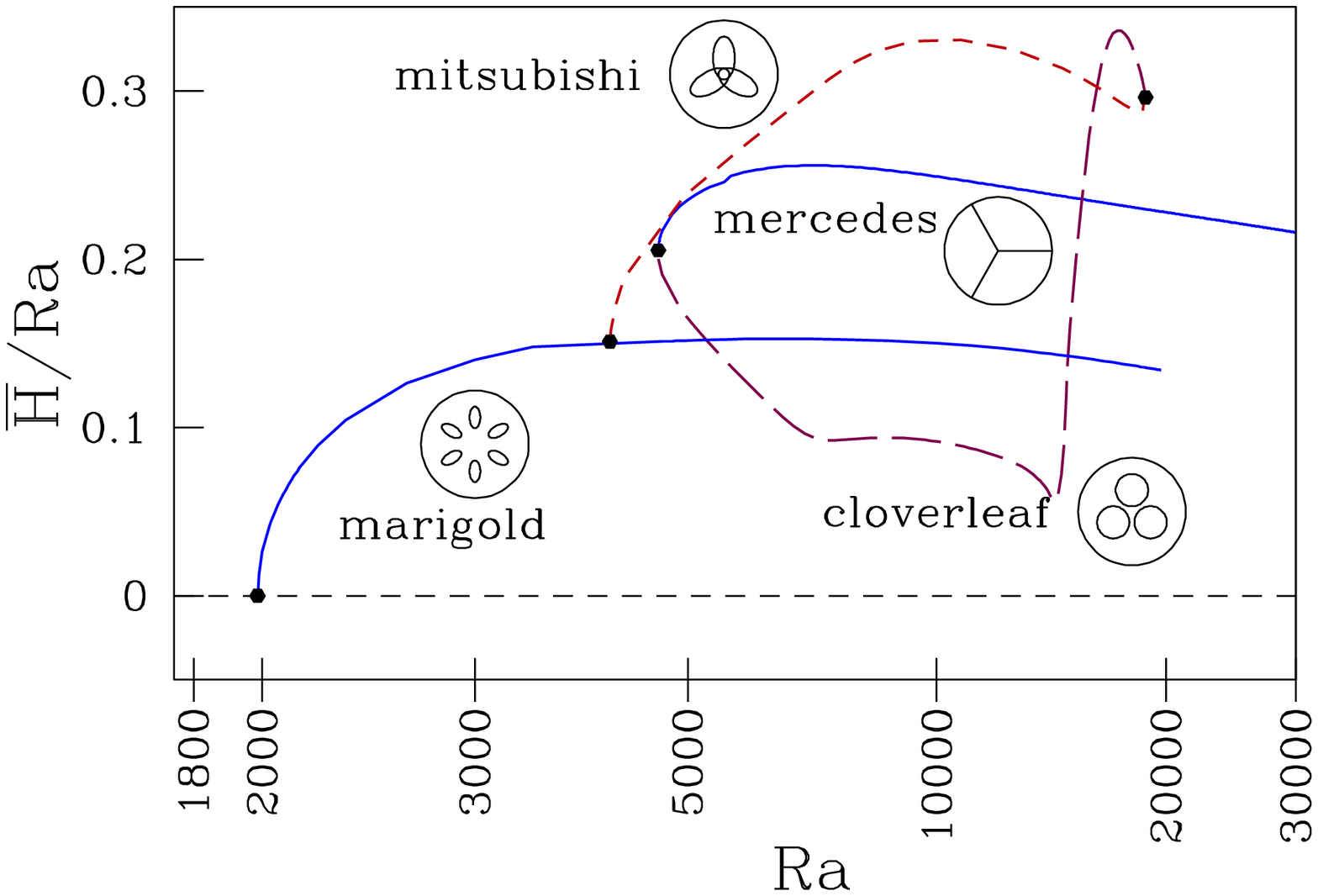} 
\caption{(Color online) Partial bifurcation diagram including only branches with three-fold
  symmetry.  The {\bf Marigold} branch (solid, blue) arises at a pitchfork
  bifurcation from the {\bf Conductive} branch (short-dashed, black) at
  $Ra=1985$.  A secondary pitchfork bifurcation from the marigold branch at
  $Ra=4103$ gives rise to the {\bf Mitsubishi} branch (short-dashed, lighter
  purple).  At a turning point at $Ra=18\,762$, it meets the 
{\bf Cloverleaf} branch (long-dashed, darker purple).
The {\bf Mercedes} branch (solid, blue) originates at another 
turning point at $Ra=4634$.}
\label{fig:merc_sm}
\includegraphics[width=12cm]{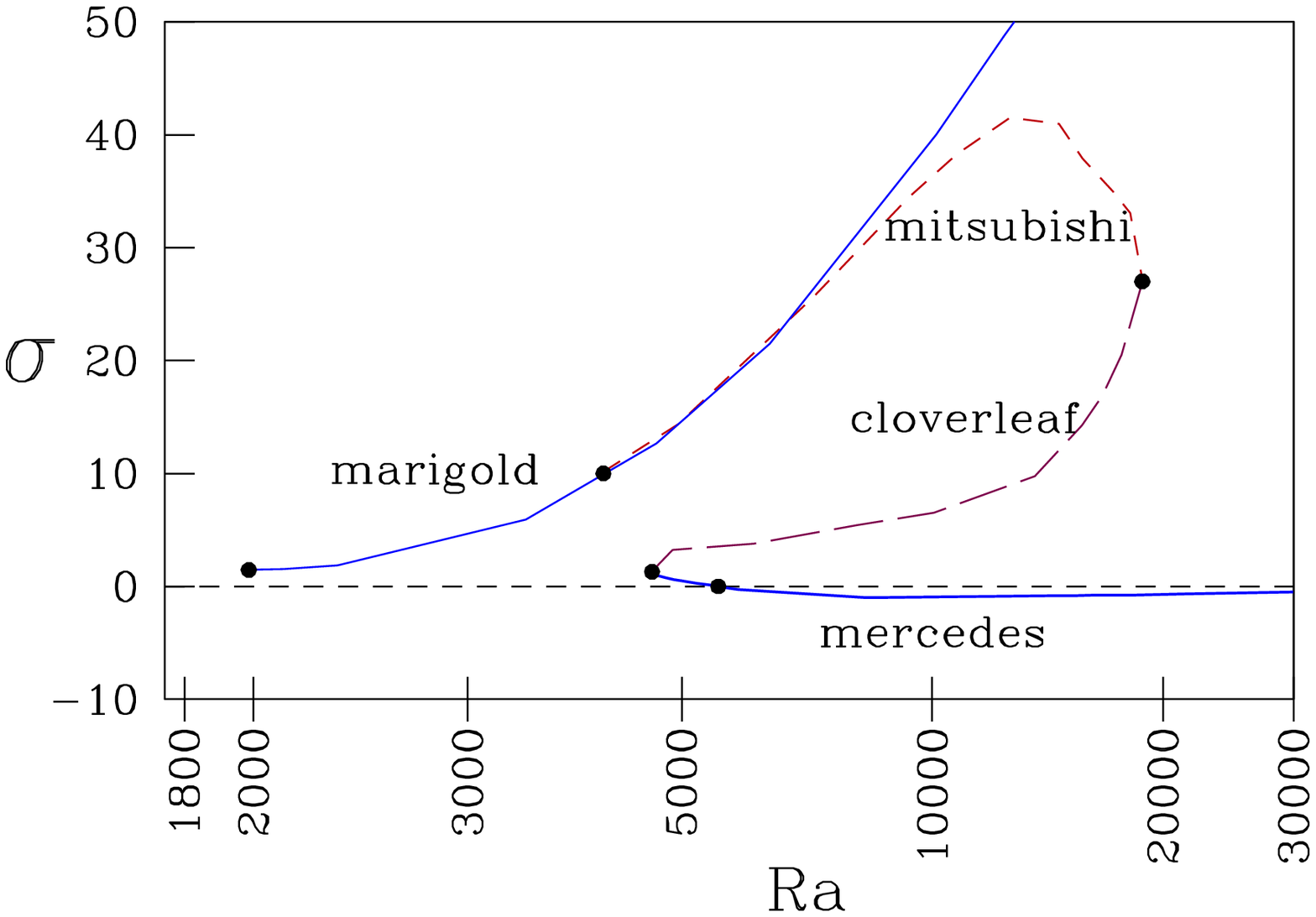}
\caption{(Color online) Leading eigenvalue for each of the three-fold-symmetric branches.
  From highest to lowest: marigold (solid, blue
Mitsubishi (short-dashed, light purple), cloverleaf (long-dashed, dark purple), 
Mercedes (solid, blue).
Dots indicate bifurcations from conductive to 
marigold branch ($Ra=1985$, $\sigma\approx 1.46$),
to Mitsubishi branch ($Ra=4103$, $\sigma\approx 10$),
to cloverleaf branch, ($Ra=18\,762$, $\sigma\approx 27$),
to Mercedes branch ($Ra=4634$, $\sigma\approx 1.3$), 
and final stabilization of Mercedes branch ($Ra=5503$, $\sigma=0$).}
\label{fig:merc_eigs}
\end{figure}

\begin{figure}[htp]
\begin{center}
\includegraphics[width=17cm]{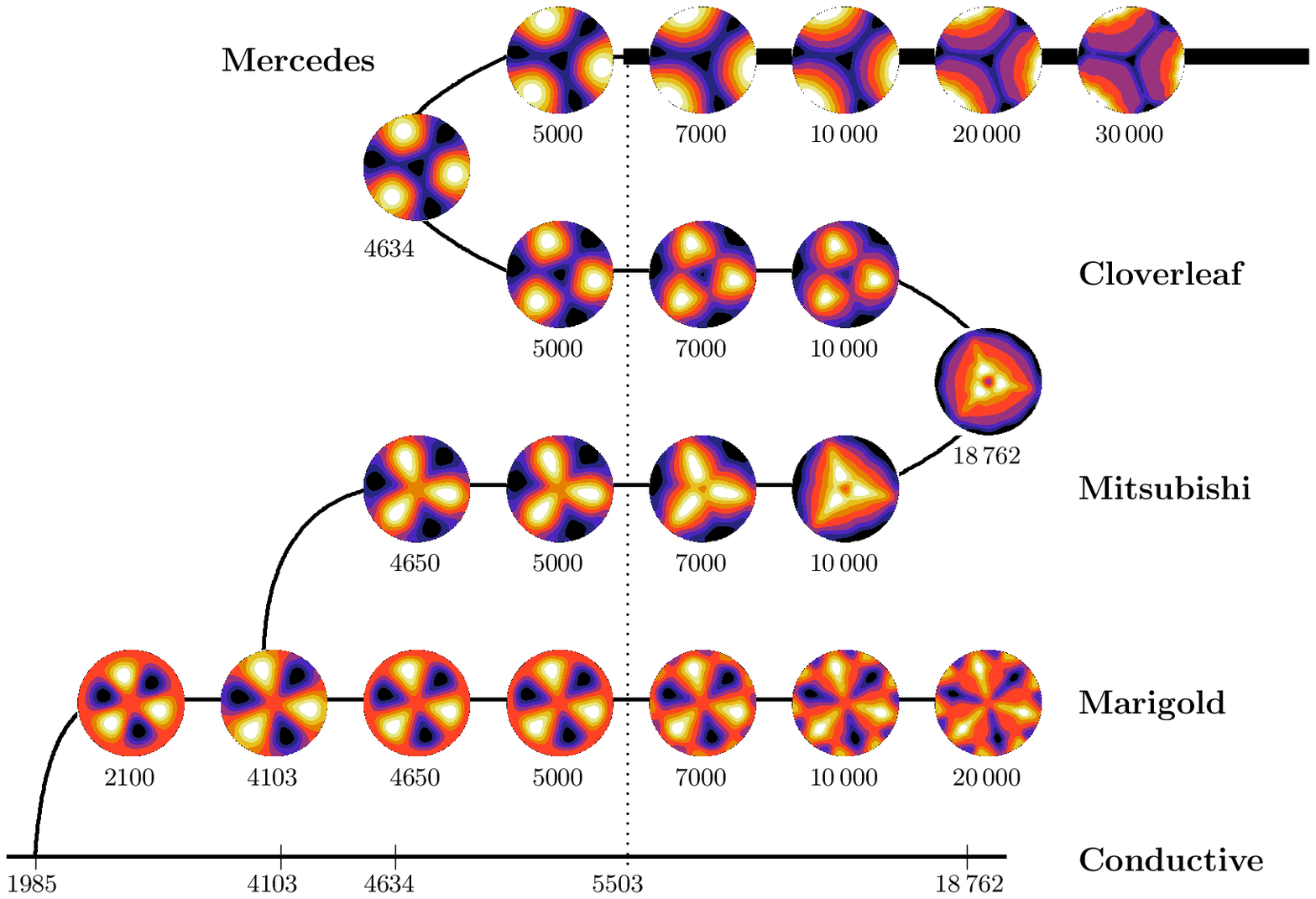}
\end{center}
\caption{(Color online) Schematic partial bifurcation diagram relating branches 
originating from the $m=3$ bifurcation.
The four branches of steady states with three-fold symmetry are 
the {\bf Mercedes}, {\bf Cloverleaf}, {\bf Mitsubishi}, and 
{\bf Marigold} branches.  The Marigold branch is created
by an $m=3$ circle pitchfork bifurcation from the Conductive branch 
at $Ra=1985$. It undergoes a pitchfork bifurcation at $Ra=4103$,
leading to the Mitsubishi branch. A turning point at $Ra=18\,762$ leads 
to the cloverleaf branch, and another turning point at 
$Ra=4634$ 
to the Mercedes branch. Only the Mercedes branch is stable, 
for $Ra > 5503$, as indicated by the thick line.}
\label{fig:merc_pic}
\end{figure}

The set of branches with three-fold symmetry are perhaps the most interesting,
and certainly the most aesthetic. We have been able to trace the tortuous
connection between the states obtained by time-integration (and hence
necessarily stable) and the $m=3$ primary branch (which, occurring 
after three
other primary bifurcations, is necessarily unstable).  Figure
\ref{fig:merc_sm} extracts from figure \ref{fig:bifdiag_full} the branches
with three-fold symmetry. Four branches are present (in addition to the
conductive branch), connected by saddle-node and pitchfork bifurcations, shown
as dots.  Figure \ref{fig:merc_eigs} plots the leading eigenvalues of these
four branches.  The qualitative bifurcation diagram in figure
\ref{fig:merc_pic} provides a clearer picture of the bifurcations, without the
crossings present in figure \ref{fig:merc_sm}.  We recall that an identical
set of branches, with hot and cold reversed (along with upwelling and
downwelling) also exists, and that the azimuthal orientation is arbitrary.

We begin by describing the states in figure \ref{fig:merc_pic}, beginning from
the stable state at $Ra\approx 30\,000$, which Hof called {\bf Mercedes}
because of its resemblance to the logo of this automobile~\cite{HofThesis}.
This state was observed in Hof's experiment~\cite{Hof} and in time-dependent
simulations~\cite{Boronska_PhD,Boronska_PRE1}, where it was computed for $Ra$
between 5000 and $29\,000$, as shown in figure \ref{fig:dns:neu}.  A mercedes
state was also computed at $Ra=31\,250$ by Leong~\cite{Leong} and at 
$Ra=14\,200$ by Ma \etal~\cite{Ma}.  For high $Ra$, the hot upwelling and cold
downwelling regions in the midplane are narrow,
confined to three hot spots along the lateral boundary and a central cold
Y-shaped region.  With decreasing $Ra$, the upwelling and downwelling regions
come to occupy an increasing portion of the midplane.  By the turning point at
$Ra=4634$, the three hot spots have widened, becoming almost
circular, and the center and extremities of the cold Y have widened into four
triangles. Emerging from this turning point is what we have called the 
{\bf Cloverleaf} branch. Following this branch towards
increasing values of $Ra$, the three hot spots move inwards from the boundary
and the cold Y-shaped region breaks, leaving four separate triangles. The hot
spots and the central cold triangle become smaller, while the three remaining
cold triangles narrow and cling to the lateral boundary. By the turning point
at $Ra=18\,762$, the hot spots have merged into one central triangular region,
and the cold regions form a ring occupying almost the entire circumference.
As we follow the new branch with decreasing $Ra$, the points of the triangle
expand and separate, forming oval petals or blades, while the exterior ring
forms three exterior triangles.  Midway along this branch, the states resemble
the logo of the {\bf Mitsubishi} automobile, and this is the name we have
given to the branch.  The hot uprising and cold downwelling regions become
more similar as $Ra$ decreases. At $Ra=4103$, the Mitsubishi branch is 
seen to emanate in a pitchfork bifurcation from the {\bf Marigold} branch, 
whose states have six equal petal-shaped regions. The marigold branch itself
is generated at a circle pitchfork bifurcation from the conductive branch at
$Ra=1985$.

The Mitsubishi, cloverleaf and Mercedes states have $D_3$ symmetry, while the
marigold states have the larger symmetry group $D_3\times Z_2$.

The cloverleaf and Mitsubishi branches were obtained from the Mercedes branch
by going around the turning points via the quadratic extrapolation described
in section \ref{sec:numerical}. Additional effort is required to 
switch from the Mitsubishi to the marigold branch, since straigtforward 
continuation treats the pitchfork as it would a turning point.
Because we can calculate eigenvectors, steady
states and transient behavior, there are a number of ways in which a
starting point on the marigold branch could be obtained: \\ (i) Take a
Mitsubishi state, for which the left and right-hand-sides of \eqref{eq:dmz}
are not equal, and average the two expressions.\\ (ii) Add a small amount of
the $m=3$ eigenvector to the conductive branch.\\ (iii) Carry out
time-integration for $Ra>1985$, retaining only azimuthal modes which are
multiples of three.\\ We used method (iii), halting the integration after the
marigold state was reached but before its instability was manifested.

Although figure \ref{fig:merc_eigs} shows leading eigenvalues corresponding to
the states of figure \ref{fig:merc_sm}, it is of a different nature from the
previous eigenvalue plots.  Figures \ref{fig:pizza_eigs} and
\ref{fig:axi_eigs} showed one or more leading eigenvalues for states along a
single branch.  In contrast, figure \ref{fig:merc_eigs} shows a single
eigenvalue per state, but for states along the four different branches
described above. Thus, between one and four eigenvalues are shown for a single
Rayleigh number.  All of the branches have at least one positive eigenvalue
(and are thus unstable) except the Mercedes branch for $Ra > 5503$.  When the
marigold branch bifurcates from the conductive branch at $Ra=1985$, it
inherits three positive eigenvalues, the largest of which is the highest curve
in figure \ref{fig:merc_eigs}.  The Mitsubishi branch shares the spectrum of
the marigold branch at the pitchfork bifurcation at $Ra=4103$.  As $Ra$
increases, both branches become more unstable, but, for $Ra \gtrsim 12\,500$,
the eigenvalues of the Mitsubishi branch eventually begin to decrease. The
Mitsubishi branch is still unstable when it meets the cloverleaf branch at the
turning point at $Ra=18\,762$.  Following the cloverleaf branch with
decreasing $Ra$, the leading eigenvalue decreases.  When the cloverleaf and
Mercedes branches meet at the turning point at $Ra=4634$, the leading
eigenvalue is still barely positive. Following the Mercedes branch with
increasing $Ra$, the leading eigenvalue continues to decrease, becoming
negative and stabilizing the branch for $Ra > 5503$.

\subsection{Dipole, Tiger and Three-roll branches ($\mathbf{m=1}$)}
\label{sec:dipole}

\begin{figure}[htp]
\includegraphics[width=12cm]{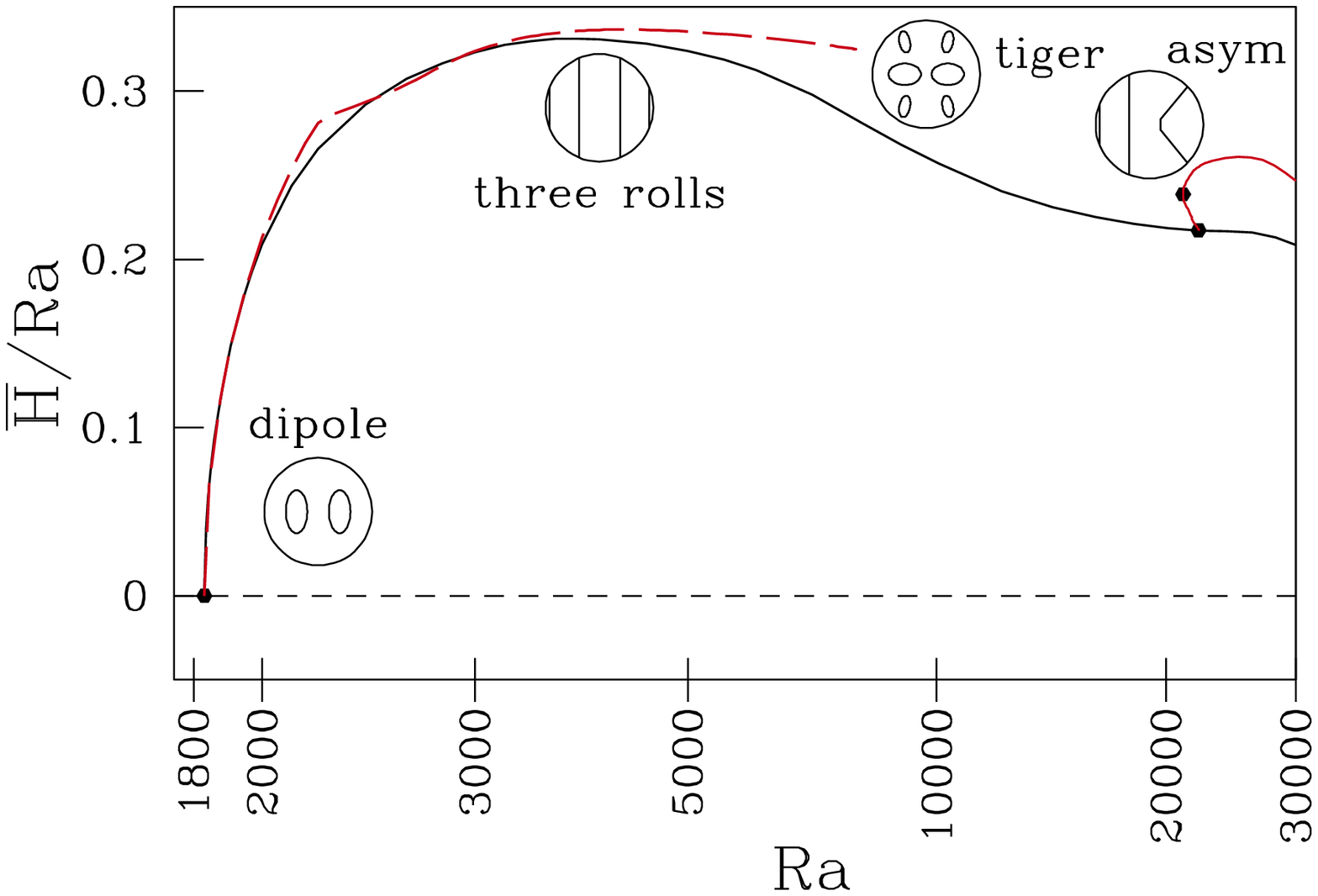}
\caption{(Color online) Partial bifurcation diagram of branches
created at the $m=1$ primary bifurcation.
Two branches of dipole states are created at $Ra=1828$,
and seem to be tangent to one another near the bifurcation.
As $Ra$ increases, the spatial forms along these branches 
evolve in divergent ways. One branch contains the three-roll states 
(solid, black) and the other (dashed, brick) contains 
the tiger states.
An asymmetric three-roll branch (solid, brick) is 
created at a subcritical pitchfork bifurcation at $Ra=22\,155$
and reverses direction at a saddle-node bifurcation at $Ra=21\,078$.}
\label{fig:dipole_sm}
\vspace*{1cm}
\begin{center}
\includegraphics[width=17cm]{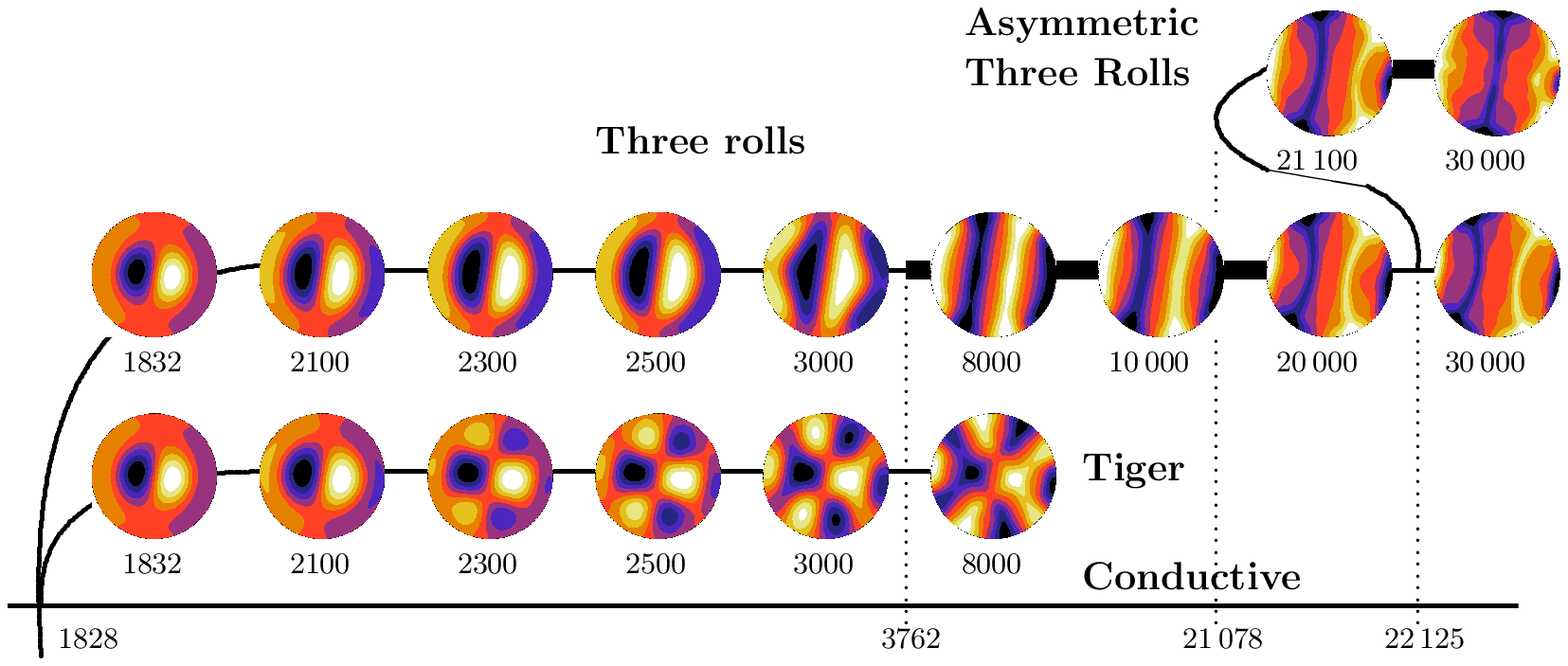}
\end{center}
\caption{(Color online) Schematic bifurcation diagram of branches emanating from $m=1$
  instability. 
Two branches bifurcate simultaneously at $Ra=1828$.
Near the bifurcation, states along both branches have the form of a dipole.
States along one branch evolve into a {\bf Three-roll} state, while
  along the other branch evolve to a form called {\bf Tiger}.   
The three-roll state is stable for $3762\leq Ra \leq 20\,393$, 
and the tiger branch is never stable.  The {\bf Asymmetric
    Three-Roll} branch is formed at $Ra=22\,125$, via a subcritical pitchfork
  bifurcation from the three-roll branch, reversing direction at a saddle-node
  bifurcation at $Ra=21\,078$.}
\label{fig:dipole_pic}
\end{figure}
\begin{figure}
\begin{center}
\includegraphics[width=17cm]{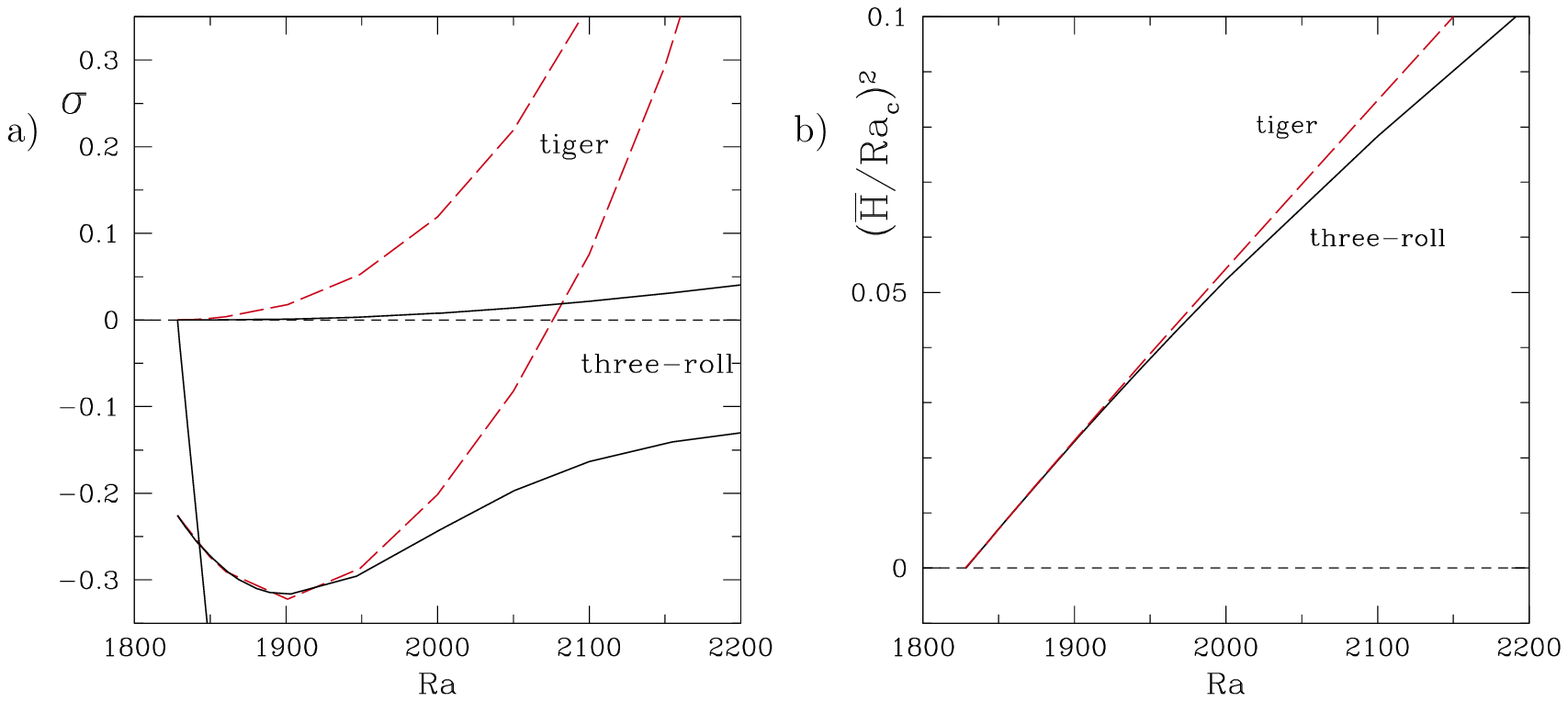}
\end{center}
\caption{(Color online)  a) Leading eigenvalues of the 
tiger (long-dashed, brick) and
three-roll (solid, black) 
  branches. The eigenvalues of the tiger branch grow rapidly as $Ra$
  increases, while the three-roll branch is weakly unstable until $Ra=3762$.
The rapidly decreasing eigenvalue is that associated with the formation 
of these branches.
b) The tiger and three-roll branches have the same curvature near onset.}
\label{fig:dipole_eigs}
\end{figure}

Although the $m=1$ bifurcation has the lowest Rayleigh number threshold,
$Ra=1828$, we have postponed its discussion because of its odd behavior.  Two
branches bifurcate simultaneously, as shown in figures
\ref{fig:dipole_sm} and \ref{fig:dipole_pic}.  The states close to the
bifurcation are of dipole form, as expected.  A dipole state is observed at
$Ra=2500$ by Leong~\cite{Leong} and by Ma \etal~\cite{Ma}.  Along one branch,
additional spots appear on either side of the dipole, which grow as $Ra$ is
increased; we have given the name of {\bf Tiger} to this branch.
We have been unable to
compute the tiger branch above $Ra=7936$.  The other branch is more
conventional. The two parts of the dipole elongate along the dipole axis and
patches of opposite sign appear and elongate near the boundary, leading
eventually to a {\bf Three-roll} structure.  The three-roll branch exists at
least until $Ra=30 \,000$, where we have stopped our computations.

Time-dependent simulations~\cite{Boronska_PhD,Boronska_PRE1} between
$Ra=20\,000$ and $25\,000$ show a transition to an {\bf Asymmetric Three-roll}
state, for which the rolls are slightly shifted. We have determined that an
asymmetric branch emerges from the three-roll branch at $Ra=22\,125$ via a
subcritical pitchfork bifurcation (the only subcritical bifurcation we have
found in this system).  The branch reverses direction and stabilizes at a
saddle-node bifurcation at $Ra=21\,078$.  Three-roll and
asymmetric three-roll states are observed at $Ra=12\,500$ and $Ra=25\,000$,
respectively, by Leong~\cite{Leong}.

The tiger and the three-roll branches share the same spatial symmetry, 
that is $D_1\times Z_2$, where $D_1$ (equivalent to $Z_2$) 
is generated by reflection in the axis perpendicular 
to the roll or dipole axis, as in \eqref{eq:dmref},
and $Z_2$ is generated by simultaneous rotation by $\pi$ and reflections
$z\rightarrow -z$, $U_z\rightarrow -U_z$, $H\rightarrow -H$, as in \eqref{eq:dmz}.
The asymmetric three-roll branch has symmetry $D_1$.

\begin{figure}[htp]
\begin{center}
\includegraphics[width=8cm]{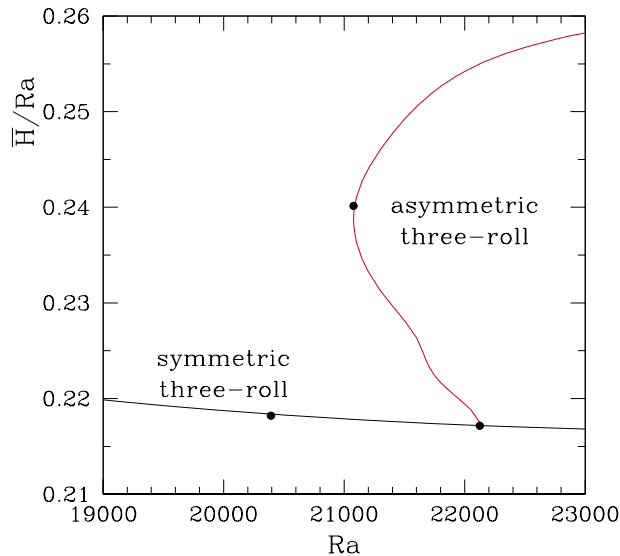}
\end{center}
\caption{(Color online) Enlargement of bifurcation diagram of figure \ref{fig:dipole_sm}.
 The symmetric three-roll branch loses stability at $Ra=20\,393$.
  A subcritical bifurcation at $Ra=22\,125$ from the symmetric three-roll
  branch leads to the creation of the asymmetric three-roll branch, which is
  stabilized by a saddle-node bifurcation at $Ra=21\,078$.}
\label{fig:asym_3r}
\end{figure}

Figure \ref{fig:dipole_eigs}\textit{a}, containing leading eigenvalues of each
branch near threshold, shows that the tiger branch becomes quite unstable
immediately after it forms.  At the same time, the three-roll branch also
becomes weakly unstable, though it eventually stabilizes at $Ra=3762$.  
Instability near onset is another unexpected feature of these branches.  
%
The three-roll branch becomes unstable again at $Ra = 20\,393$, as shown in
figure \ref{fig:asym_3r}. 

We have sought to better understand the primary $m=1$ bifurcation, at which
both the tiger and three-roll branches bifurcate simultaneously.  First, we
have verified that the three-roll and the tiger branches are distinct by
following them around the pitchfork bifurcation, producing 
symmetrically-related branches.
Thus the possibility that our continuation procedure has jumped from one
branch to another is ruled out.
Figure \ref{fig:dipole_eigs}\textit{b} shows that both the tiger and the
three-roll branches emerge via a pitchfork bifurcation, i.e.~that
$\Hmax \propto \sqrt{Ra-Ra_c}$, or, equivalently, that
\begin{equation}
\left(\frac{\Hmax}{Ra_c}\right)^2 = \alpha \frac{Ra-Ra_c}{Ra_c}
\label{eq:pf}
\end{equation}
for $Ra_c =1828.37 \leq Ra \lesssim 1900$.
Figure \ref{fig:dipole_eigs}\textit{b} also shows that the constant of
proportionality in \eqref{eq:pf} is the same for
both branches ($\alpha \approx 0.59$). Thus the two branches initially
share not only a vertical tangent, but even the same curvature.

Simultaneously bifurcating and non-equivalent branches are encountered in a
number of situations, notably pitchfork bifurcation in the presence of $D_4$
symmetry, such as in a square box~\cite{CK,BHK}.  In the $D_4$ case, one set of
branches contains solutions whose axes of symmetry are the vertical or the
horizontal midline of the square, while, for the other set of solutions, the
symmetry axes are the diagonals. Although the two types of branches bifurcate
simultaneously, they are not related to one another by a symmetry operation of
$D_4$, and so are not dynamically equivalent.
For $O(2)$, in contrast to $D_4$, the concepts of vertical, horizontal and
diagonal have no meaning: solutions of any orientation can be obtained by
rotation, and so must all be dynamically equivalent.  
An explanation of the $m=1$ behavior is the subject of a separate investigation.

Most of the results given above are consistent with the time-dependent
simulations of our companion paper~\cite{Boronska_PhD,Boronska_PRE1},
summarized in figure \ref{fig:dns:neu}. The inconsistencies can largely
explained as a consequence of finite integration time and very weak
instability.  This accounts for the observation of a symmetric three-roll
state for $25\,000\leq Ra \leq 30\,000$, a range over which it is unstable,
rather than its stable asymmetric counterpart, The same is true of the
long-lived dipole state observed near onset at $Ra=2000$.  
In all of these cases, the largest eigenvalue is less than 0.5, 
which would not have allowed initally small perturbations to grow to 
appreciable levels over the duration of our time-dependent simulations;
this is true for all of the steady states shown in figure \ref{fig:dns:neu} 
summarizing the time-dependent simulations.



Time-dependent simulation from an initial dipole
state at various Rayleigh numbers also yielded various interesting
transients~\cite{Boronska_PhD,Boronska_PRE1}, and led to the discovery of an
oscillatory pattern, the rotating S, and of two new steady patterns: the
dipole smile, whose branch we have not continued, and the CO, described in the
next section.

\subsection{Two-roll and CO branches}

\begin{figure}[!h]
\begin{center}
\includegraphics[width=12cm]{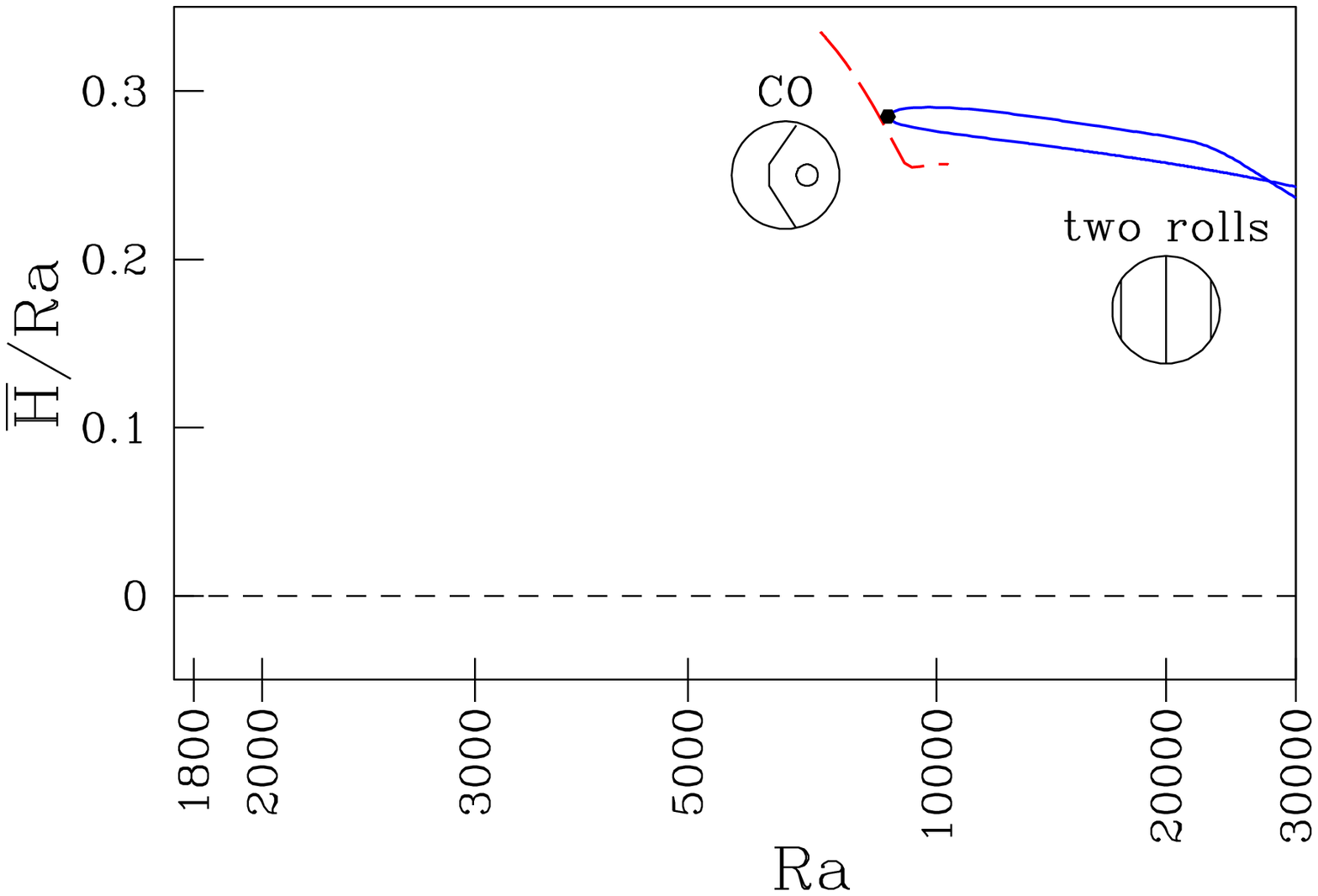}
\end{center}
\caption{(Color online) Partial bifurcation diagram containing the two-roll
and CO branches.
Two-roll branches (solid, blue) arise via saddle-node bifurcation at
$Ra=8677$. CO branch (dashed, red) has been computed for
$7167\leq Ra \leq 10\,348$.
The apparent intersections 
(between the two-roll and CO branches and between the stable and unstable 
two-roll branches) is an artifact of the projection.
}
\label{fig:tworolls_sm}
\vspace*{1cm}
\begin{center}
\includegraphics[width=12cm]{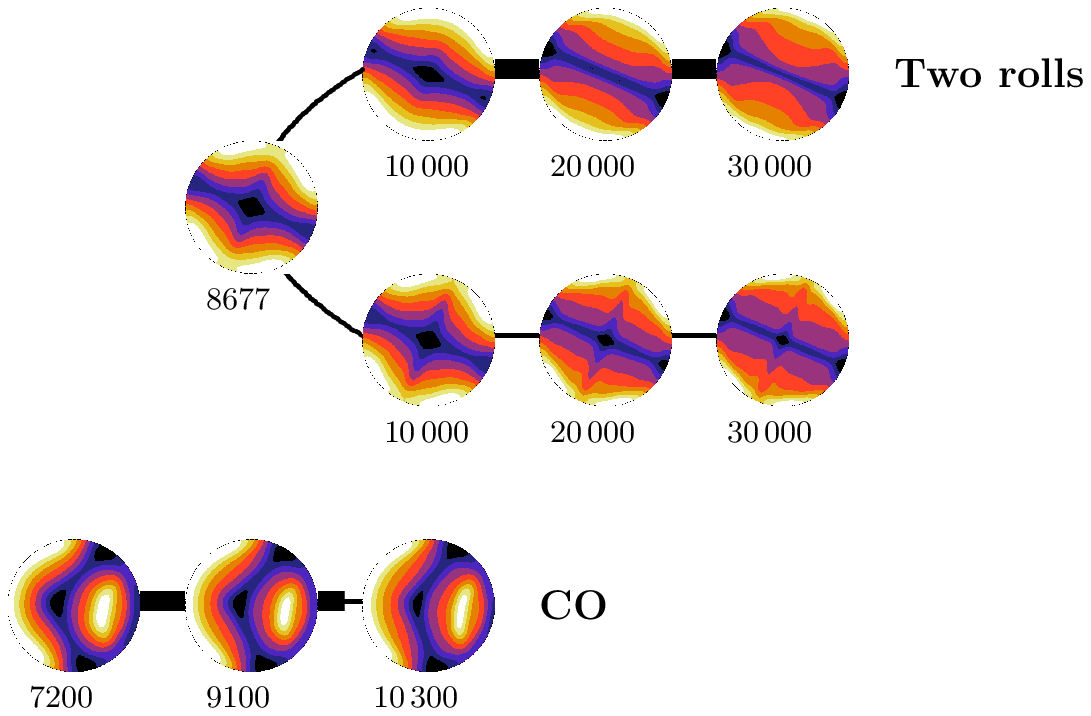}
\end{center}
\caption{(Color online) Schematic partial bifurcation diagram showing the two-roll and CO
  branches. The two-roll branches originate at a turning point at $Ra=8677$
  and have been computed for $Ra\leq 30\,000$.  The upper branch is stable for
  $Ra \leq 28\,086$.  The CO branch has been calculated in the range $7167\leq
  Ra \leq 10\,348$; it is stable for $Ra \leq 10\,087$.}
\label{fig:tworolls_pic}
\end{figure}
\begin{figure}
\begin{center}
\includegraphics[width=17cm]{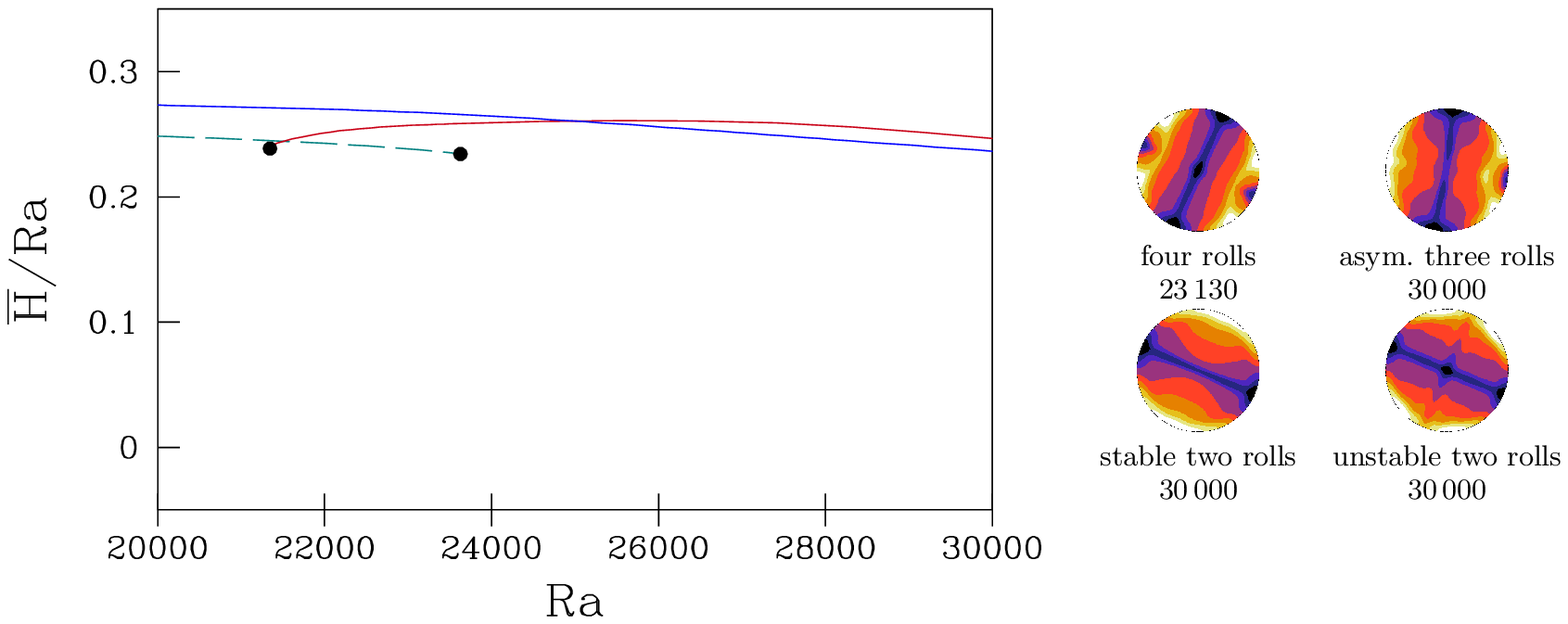}
\end{center}
\vspace*{-1cm}
\caption{(Color online) Four-roll (dashed, turquoise), asymmetric three-roll (solid, brick) 
and two-roll branches (solid, blue) at high $Ra$.}
\label{fig:bifhigh}
\end{figure}

Finally, we mention another set of branches which, like the one-torus
branches, appear to be unconnected to the conductive state.  These are the
{\bf Two-roll} and the {\bf CO branch}, shown in figures \ref{fig:tworolls_sm}
and \ref{fig:tworolls_pic}.  

Figure~\ref{fig:dns:neu} shows how these branches were originally
found by time-integration, staring from quasi-steady states at $Ra=2000$.
A CO state was found by starting from a dipole and setting $Ra=10\,000$ and a
two-roll state by starting from a pizza and setting $Ra=16\,000$.
The two-roll branch originates at a saddle-node bifurcation at $Ra=8677$,
where it is connected to an unstable branch containing states which 
also have two rolls. Figure~\ref{fig:tworolls_pic} shows that, for 
high $Ra$, states on the stable branch have an indentation in the 
central boundary which divides the rolls, while those on the unstable 
branch have a protrusion. 
Two-roll states have also been observed in the
numerical simulations at $Ra=14\,200$ by Ma \etal~\cite{Ma} and at
$Ra=37\,500$ by Leong~\cite{Leong}.
The two-roll branches are very robust:
both exist at least until $Ra=30\,000$ and the upper branch is 
stable for $Ra\leq 28\,086$. We have been able to compute the CO
branch only for $7167 \leq Ra \leq 10\,348$, and it is stable for
$Ra \leq 10087$.
The two-roll states have two symmetry axes (i.e.~$D_2$), while the CO states,
containing two light regions, one curved and one oval) have only one symmetry
axis ($Z_2$).

Figure \ref{fig:bifhigh} shows that the high Rayleigh number ($Ra \geq 20\,000$)
asymmetric three-roll states and four-roll states 
greatly resemble two-roll states.  Along all of the branches, the convective
structures widen as $Ra$ increases for all the branches: this is the form
taken in this confined geometry of the well-known increase in wavelength for
large systems of parallel rolls.  For the branches emerging from the $m=1$ and
$m=2$ primary bifurcations, this tendency eventually leads, after secondary
bifurcations and more gradual deformations, to states which primarily contain
two rolls.  These are far removed from the trigonometric forms of the dipole
and pizza states that prevail along these branches at low $Ra$.

\section{Conclusion}
\label{sec:conclusion}

\begin{figure}
\begin{center}
\includegraphics[width=18cm]{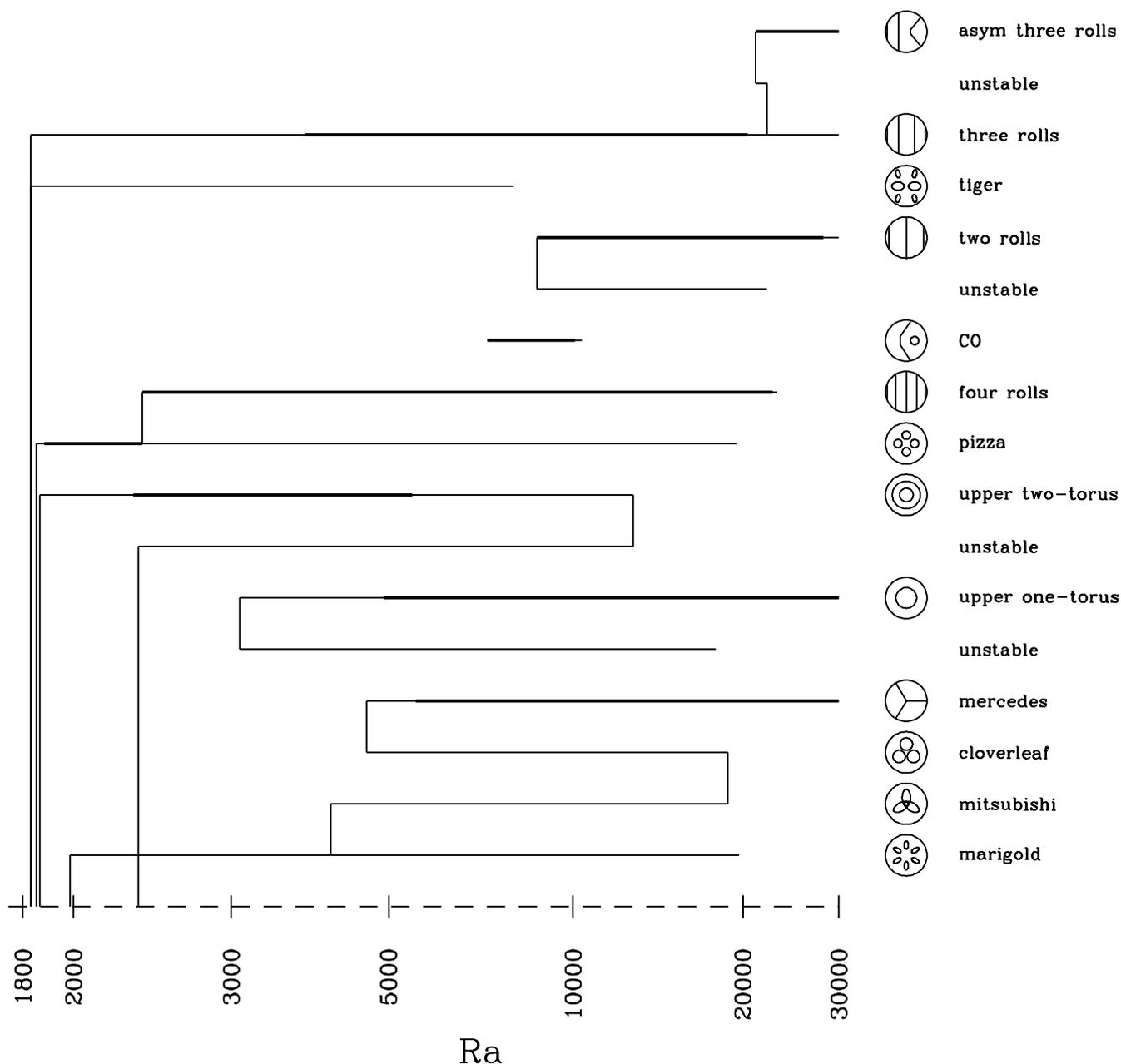}
\end{center}
\caption{Schematic bifurcation diagram. Arbitrary quantity of the vertical
  axis chosen to eliminate all but one intersection.  Bold lines indicate
  stable portions of branches.}
\label{fig:bifsch}
\end{figure}

We have presented an intricate bifurcation diagram describing
Rayleigh-B\'enard convection in a cylinder with aspect ratio $\Gamma=2$ and
$Pr=6.7$ for $Ra\leq 30\,000$.  This study is complementary to the
time-dependent simulations described in our companion paper
\cite{Boronska_PRE1}; the branches of the bifurcation diagram were obtained by
continuation from the stable states resulting from time integration.  We have
determined the bifurcation-theoretic origin of these states, including the
five states observed experimentally by Hof \etal.  In one case, the path is
straightforward: the four-roll branch results from a secondary bifurcation
from the pizza branch, which in turn arises from a primary $m=2$ bifurcation
from the conductive branch.  In another case, it is more tortuous: for $m=3$,
two additional saddle-node bifurcations must be traversed between the primary
marigold branch and the stable Mercedes branch which is actually observed.
The torus and two-roll branches turned out to be disconnected (as far as we
can tell) to the conductive state.  Finally, we have located the primary $m=1$
bifurcation leading to the three-roll branch, but it is accompanied by another
simultaneously bifurcating branch and has unexpected stability properties. We
have also traced the disconnected CO branch
and the two-torus branch arising from a primary $m=0$ bifurcation.
A schematic version of the bifurcation diagram is given in 
figure \ref{fig:bifsch}, while tables \ref{tab:allbranches} and 
\ref{tab:allbifs} list all of the branches 
we have obtained, as well as the bifurcations and their nature.

The diagram we have obtained contains 17 branches of steady states, but is
nonetheless incomplete. Although we have followed the primary branches
originating at 5 bifurcations along the conductive branch, there are literally
hundreds of other primary bifurcation points in the range $Ra\leq 30\,000$.
Each of the primary branches thus engendered can and does undergo many
secondary bifurcations.  In addition, while calculating the stability of the
various branches we have observed many eigenvalues cross zero,
signalling the appearance of a new branch.  Finally, there is no way to
ascertain how many other disconnected branches. 
It is surely unfeasible and unproductive to
strive to find all branches.

Despite the complexity of the bifurcation diagram, its main features can be
described quite simply.  Circle pitchfork bifurcations to trigonometric
branches dictated by the geometry -- dipole, pizza, marigold, two-torus --
take place at low Rayleigh numbers. These undergo various secondary
bifurcations at intermediate Rayleigh numbers that lead to states with
rolls. At high Rayleigh numbers, there are three types of stable branches:
torus, Mercedes, and states essentially containing two rolls.

Our goal of relating the states obtained by time integration to the
bifurcation diagram is largely realized, but there remain a few loose ends.
One is to complete our understanding of the two simultaneously bifurcating
$m=1$ branches.
Another small and clear-cut goal is to incorporate the time-periodic rotating
S state we have described in~\cite{Boronska_PhD,Boronska_PRE1} into the
bifurcation diagram by ascertaining its bifurcation-theoretic origin and exact
domain of existence and stability.  Comparing our study to that of Ma et
al., we find very similar values for the five primary pitchfork
bifurcations, for the secondary pitchfork bifurcation creating the four-roll
branch, and for the two secondary pitchfork bifurcations which stabilize the
two-torus branch.  The other bifurcations in Table \ref{tab:allbifs} are not
present in Ma \etal. These authors did, however, compute an additional
stable four-spoked pattern at $Ra = 14\,200$, whose
bifurcation-theoretic genesis could be interesting to determine.

Our study also points in several larger directions.  It would be desirable to
incorporate improved versions of the numerical methods we have used, namely
adjustment of $Ra$ within the Newton iteration, the inverse Arnoldi method,
and the calculation of traveling waves as steady states in a rotating frame.
This example could also be used as a test case to try to understand and to
control the enormous variability in performance of BiCGSTAB in solving the
linear equations of Newton's method in different regions of the bifurcation
diagram.  Finally, an extensive but straightforward goal is to compute a
bifurcation diagram for the case in which the sidewalls are thermally
conducting rather than insulating.  Time-dependent simulations have already
yielded initial conditions and approximate stability ranges for many
branches~\cite{Boronska_PhD,Boronska_PRE1}. For the conducting case, as for
the insulating case, construction of a complete bifurcation diagram is
doubtless impossible, but the stable steady and time-periodic states could all
be traced back to their bifurcation-theoretic origin from the conductive state
as we have done here.

The complexity of the bifurcation diagram we have computed is interesting in
light of the recent computational discovery of large numbers of unstable
solutions of wall-bounded shear flows, e.g.~\cite{Pipe}.  It is hypothesized
that weak turbulence can be understood as chaotic trajectories,
e.g.~\cite{Predrag_Visualizing}, that visit in turn the vicinities of the
various unstable branches, which are created at saddle-node bifurcations.  Our
study contributes two observations to this line of research.  First, this
example provides a reminder that the existence of a large number of unstable
solutions is a typical property of the hydrodynamic equations.  Second, our
study underlines the fact that such multiplicity can occur in the absence of
complicated dynamics.

This study showcases our numerical methods for carrying out time-integration,
branch continuation and linear stability analysis by using a single code with
several different high-level drivers.  Newton's method has many advantages: it
is much faster and much more precise, and can compute unstable states. Time
integration remains, nevertheless, absolutely essential for generating initial
states, especially since several important branches are disconnected from the
conductive state.  Although our cylindrical Rayleigh-B\'enard computation is
quite specific, it demonstrates what can be accomplished for three-dimensional
nonlinear problems by combining matrix-free preconditioned numerical methods
with dynamical systems theory.

\begin{table}[htp]
\begin{center}
\includegraphics[width=11cm]{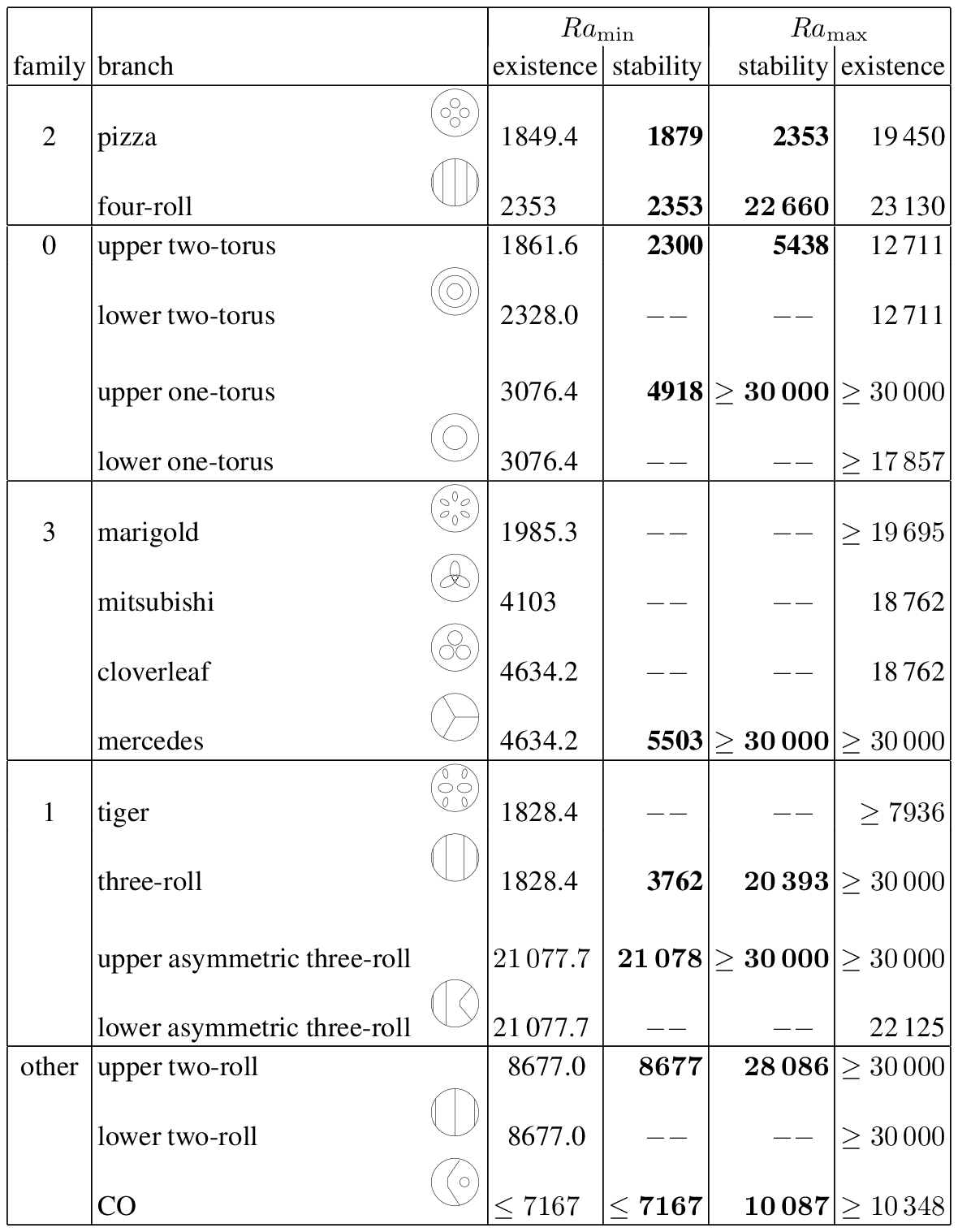}
\end{center}
\caption{List of all convective branches computed, with
lower and upper limits of existence and stability.
Stable patterns (between limits in boldface) should be observable 
in experiments or time-dependent simulations.
Inequalities ($\geq$ or $\leq$) indicate a lower or upper bound for
the corresponding $Ra$:
calculation of the branch either terminated for an unknown reason,
or continued past 30\,000.
Dashes ($--$) indicate the lack of stability limits for branches 
which are unstable throughout.}
\label{tab:allbranches}
\end{table}

\begin{table}[htp]
\begin{center}
\begin{tabular}{|l|l|l|l|}
\hline
family & bifurcation & ~$Ra$ & comments \\\hline
2 & circle pitchfork & ~1849.4 & creates pizza branch\\
  & pitchfork & ~2353 & creates four-roll branch\\
  & eigenvalue crossing & ~1879 & stabilizes pizza branch \\
  & turning point & $19\,450$ & terminates pizza branch \\
  & eigenvalue crossing & $22\,660$ & destabilizes four-roll branch \\
  & turning point & $23\,060$ & terminates four-roll branch \\\hline
0 & pitchfork & ~1861.6 & creates upper two-tori branch \\
  & pitchfork & ~2328.0 & creates lower two-tori branch \\
  & turning point & $12\,711$ & terminates upper and lower two-tori branches \\
  & eigenvalue crossing & ~2116 & stabilizes upper two-tori branch against
$m=2$ eigenvector \\
  & eigenvalue crossing & ~2300 & stabilizes upper two-tori branch against
$m=1$ eigenvector \\
  & eigenvalue crossing & ~5438 & destabilizes upper two-tori branch against
$m=1$ eigenvector \\\hline
0  & turning point & ~3076.4 & creates upper and lower one-torus branches \\
  & eigenvalue crossing & ~3330 & 
stabilizes upper one-torus branch against $m=6$ eigenvector \\
  & eigenvalue crossing & ~3438 & 
stabilizes upper one-torus branch against $m=2$ eigenvector \\
  & eigenvalue crossing & ~4408 & 
stabilizes upper one-torus branch against $m=5$ eigenvector \\
  & eigenvalue crossing & ~4582 & 
stabilizes upper one-torus branch against $m=3$ eigenvector \\
  & eigenvalue crossing & ~4918 & 
stabilizes upper one-torus branch against $m=4$ eigenvector \\\hline
3 & circle pitchfork & ~1985.3 & creates marigold branch \\
  & pitchfork & ~4103 & creates mitsubishi branch \\
  & turning point & ~4634.2 & creates mercedes and cloverleaf branches \\
  & turning point & $18\,762$ & terminates cloverleaf and mitsubishi branches \\
  & eigenvalue crossing & ~5503 & stabilizes mercedes branch \\\hline
1 & circle pitchfork & ~1828.4 & creates tiger and three-roll branches \\
  & eigenvalue crossing & ~3762 & stabilizes three-roll branch\\
  & eigenvalue crossing & $20\,393$ & destabilizes three-roll branch\\
  & turning point & 21\,077.7 & creates asymmetric three-roll branch \\
  & subcritical pitchfork & $22\,125$ & terminates asymmetric three-roll branch \\\hline
other & turning point & ~8677.0 & creates two-roll branches \\
      & eigenvalue crossing & $28\,086$ & destabilizes two-roll branch\\
      & eigenvalue crossing & $10\,087$ & stabilizes CO branch\\\hline
\end{tabular}
\end{center}
\caption{List of all bifurcations located.
Circle pitchfork bifurcation breaks axisymmetry, creating ``circle'' 
of new states parametrized by azimuthal phase. Pitchfork bifurcation 
breaks a reflection symmetry, creating two branches.
Eigenvalue crossings are necessarily accompanied by bifurcations,
whose nature we have not investigated.
Comments such as ``stabilizes'', ``destabilizes'',``creates'', ``terminates'' 
are to be interpreted in the direction of increasing Rayleigh number.}
\label{tab:allbifs}
\end{table}

\begin{acknowledgments}

We thank Tom Mullin and Bj\"orn Hof for their continued interest in this work.
We thank Dwight Barkley, Edgar Knobloch, Laurent Martin-Witkowski, Paul
Matthews, and Isabel Mercader for useful discussions. All of the computations
were performed on the computers of IDRIS (Institut pour le Developpement des
Ressources Informatiques et Scientifiques) of CNRS (Centre National pour la
Recherche Scientifique) under project 1119.  

\end{acknowledgments}


\clearpage
\begin{thebibliography}{99}

\bibitem{Hof}
B.~Hof, G.J.~Lucas \& T.~Mullin,
{\sl Flow state multiplicity in convection},
Phys.~Fluids {\bf 11}, 2815--2817 (1999).

\bibitem{HofThesis}
B.~Hof, {\sl The visualisation of convective flow patterns in water in
  tilted cells with variable geometry,} 
Master's thesis, University of Manchester, 1997.

\bibitem{Boronska_PhD}
K.~Boro\'nska, 
{\sl Motifs tridimensionels dans la convection de Rayleigh-B\'enard cylindrique}, 
Th\`ese de doctorat, Universit\'e de Paris 7, 2005.

\bibitem{Boronska_PRE1}
K.~Boro\'nska \& L.S.~Tuckerman,
{\sl Extreme multiplicity in cylindrical Rayleigh-B\'enard convection,
I.~Time-dependence and oscillations},
Phys.~Rev.~E, submitted.

\bibitem{Benjamin}
T.B.~Benjamin \& T.~Mullin,
{\sl Notes on the multiplicity of flows in the Taylor experiment},
J.~Fluid Mech. {\bf 121}, 219--230 (1982).

\bibitem[Leong(2002)]{Leong}
S.S.~Leong, 
{\sl Numerical study of {R}ayleigh-{B}{\'e}nard convection in a cylinder}, 
Numerical Heat Transfer, Part A {\bf 41}, 673--683 (2002).

\bibitem{Ma}
D.J.~Ma, D.J.~Sun \& X.Y.~Yin,
{\sl Multiplicity of steady states in cylindrical Rayleigh-B\'enard convection}, 
Phys.~Rev.~E {\bf 74}, 03\,7302 (2006). 


\bibitem{Mamun}
C.K.~Mamun \& L.S.~Tuckerman,
{\sl Asymmetry and Hopf bifurcation in spherical Couette flow},
Phys.~Fluids {\bf 7}, 80--91 (1995).

\bibitem{Timesteppers}
L.S.~Tuckerman \& D.~Barkley,
{\sl Bifurcation analysis for time-steppers}, in 
Numerical Methods for Bifurcation Problems and Large-Scale Dynamical Systems,
ed.~by E.~Doedel \& L.S.~Tuckerman (Springer, New York, 2000), p.~452--466.

\bibitem{Vandervorst}
H.A.~van der Vorst,
{\sl Bi-CGSTAB: A fast and smoothly converging variant of
Bi-CG for the solution of nonsymmetric linear systems},
SIAM J.~Sci.~Stat.~Comput.~{\bf 13}, 631 (1992).

\bibitem{Arnoldi}
W.E.~Arnoldi,
{\sl The principle of minimized iterations in the solution of
the matrix eigenvalue problem},
Q.~Appl.~Math {\bf 9}, 17 (1951).

\bibitem{Xin}
S.~Xin, P.~Le Qu\'er\'e \& L.S.~Tuckerman,
{\sl Bifurcation analysis of double-diffusive convection
with opposing horizontal thermal and solutal gradients},
Phys.~Fluids.~{\bf 10}, 850--858 (1998).

\bibitem{Laurent}
L.~Martin-Witkowski, private communication.

\bibitem{CK}
J.D.~Crawford, E.~Knobloch, {\sl Symmetry and symmetry-breaking bifurcations 
in fluid dynamics}, Annu.~Rev.~Fluid Mech.~{\bf 23} 341 (1991).

\bibitem{BHK}
A.~Bergeon, D.~Henry \& E.~Knobloch, 
{\sl Three-dimensional Marangoni-Benard flows in square and nearly-square 
containers}, Phys.~Fluids {\bf 13}, 92--98 (2001).







\bibitem{Pipe}
B.~Hof, C.W.H.~van Doorne, J.~Westerweel,F.T.M.~Nieuwstadt, 
H.~Faisst, B.~Eckhardt, H.~Wedin, R.R.~Kerswell, F.~Waleffe,
{\sl Experimental Observation of Nonlinear Traveling Waves in Turbulent Pipe Flow}, 
Science {\bf 305}, 1594--1598 (2004).







\bibitem{Predrag_Visualizing}
J.~F.~Gibson, J.~Halcrow, P.~Cvitanovi\'c, 
{\sl Visualizing the geometry of state space in plane Couette flow},
J.~Fluid Mech.~{\bf 611}, 107--130 (2008).

\end{thebibliography}
\end{document}